\documentclass{aastex701}

\usepackage{amsmath}
\usepackage{subcaption}

\begin{document}

\title{\r{A}ngstr\"{o}m Exponent Calculation Using Wide-Field Stellar Photometry}

\author[orcid=0009-0001-8371-5794, sname='Negi']{Shefali Negi}
\affiliation{FZU -- Institute of Physics of the Czech Academy of Sciences, Na Slovance 1999/2, 182 21 Prague, Czech Republic}
\affiliation{Charles University, Faculty of Mathematics and Physics, Ke Karlovu 3, 121 16 Prague 2, Czech Republic}
\email[show]{negi@fzu.cz}

\author[orcid=0000-0001-8807-6162, sname='Ebr']{Jan Ebr}
\affiliation{FZU -- Institute of Physics of the Czech Academy of Sciences, Na Slovance 1999/2, 182 21 Prague, Czech Republic}
\email{ebr@fzu.cz}

\author[orcid=0000-0003-0035-651X, sname='Karpov']{Sergey Karpov}
\affiliation{FZU -- Institute of Physics of the Czech Academy of Sciences, Na Slovance 1999/2, 182 21 Prague, Czech Republic}
\email{karpov@fzu.cz}

\author[orcid=0000-0001-5341-6008, sname='Eliasek']{Ji\v{r}\'{i} Eli\'{a}\v{s}ek}
\affiliation{FZU -- Institute of Physics of the Czech Academy of Sciences, Na Slovance 1999/2, 182 21 Prague, Czech Republic}
\affiliation{Charles University, Faculty of Mathematics and Physics, Ke Karlovu 3, 121 16 Prague 2, Czech Republic}
\email{jiri.eliasek@matfyz.cuni.cz}

\begin{abstract}
The FRAM (F/Photometric Robotic Atmospheric Monitor) telescopes can measure the aerosol extinction at optical wavelengths using wide-field stellar photometry at the Cherenkov Telescope Array Observatory (CTAO) sites. The \r{A}ngstr\"{o}m exponent describes the wavelength dependence of aerosol extinction. In this work we extend the FRAM method from a single $B$-filter (440~nm) analysis using the Tycho-2 catalogue to a simultaneous three-filter analysis in $B$, $V$, and $R$ (440, 540, and 620~nm) using Gaia~DR3 synthetic photometry, which provides photometry directly in the Johnson--Cousins system without passband transformations. The exponent is derived from the vertical aerosol optical depth (VAOD) for the B--V, V--R, and B--R filter pairs at three sites over 2021--2022.

The multi-filter VAOD shows a systematic filter-dependent ordering that is inconsistent with physical expectation and produces negative exponents. It is not currently possible to determine which filter yields the correct VAOD, and none of the four processing configurations we tested resolves the ordering. We introduce an intercept correction as a practical remedy that yields physically plausible exponents.

At the two southern Paranal sites (CTAO-S0 and CTAO-S1) the exponent distributions are unimodal, with mean $\alpha \approx 0.84$ (B--V) and $0.65$ (B--R) at CTAO-S0 and $0.60$ and $0.41$ at CTAO-S1. At CTAO-N (La Palma) the distributions are bimodal, with $\alpha \approx 0.2$ (B--V) during episodic Saharan dust intrusions and $\alpha \approx 1.3$ under clean background conditions. AERONET sun-photometer measurements on the dates common to both instruments show the same two populations, providing qualitative support for the FRAM method.
\end{abstract}


\section{Introduction}

Imaging Atmospheric Cherenkov Telescopes (IACTs) detect very-high-energy 
gamma rays by observing the Cherenkov light produced by particle showers 
initiated in the Earth's atmosphere. Since the atmosphere is an integral part 
of the detection process, accurate characterisation of its optical properties 
is essential for the reconstruction of gamma-ray energies and fluxes 
\citep{Schmuckermaier2023, DominisPrester2024, Gaug2017}. In particular, aerosol particles introduce a 
wavelength- and altitude-dependent attenuation of Cherenkov light that must 
be accounted for in the data analysis.

At the Cherenkov Telescope Array Observatory (CTAO), a Raman LIDAR and the 
FRAM (F/Photometric Robotic Atmospheric Monitor) telescopes work together to 
characterise the aerosol content of the atmosphere. The LIDAR measures the 
full vertical aerosol extinction profile at two fixed wavelengths, while FRAM 
continuously measures aerosol extinction across the telescope field of view 
using wide-field stellar photometry. We distinguish the aerosol optical depth 
(AOD), the extinction along the line of sight, from the vertical aerosol 
optical depth (VAOD), its projection to the zenith, related by 
$\mathrm{VAOD} = \mathrm{AOD}/X$, where $X$ is the airmass; the results in 
this work are given as VAOD. 
Together they produce aerosol transmission hypercubes --- representations of 
aerosol extinction as a function of altitude, wavelength, and time --- used 
to correct the reconstructed gamma-ray energies \citep{DominisPrester2024}. 
The wavelength dimension of this hypercube requires knowledge of the 
\r{A}ngstr\"{o}m exponent, which describes how aerosol extinction varies 
across the Cherenkov light spectrum and must be determined to an absolute 
accuracy of better than $\pm 0.3$ \citep{DominisPrester2024}.

The FRAM method was originally developed for the $B$ filter 
(Johnson-Cousins system) with Tycho-2 \citep{Hog2000} as the reference catalogue 
\citep{Ebr2021}. In this work, we extend the analysis to the $V$ and $R$ 
filters using Gaia~DR3 synthetic photometry \citep{2023A&A...674A..33G}, enabling simultaneous VAOD measurements in three bands and the calculation of the \r{A}ngstr\"{o}m exponent from the B--V, V--R, and B--R pairs. The data \citep{FRAMdata} come from three CTAO sites: CTAO-N at the Observatorio del Roque de los Muchachos, La Palma, Spain, and CTAO-S0 and CTAO-S1, currently installed at an intermediate 
location at the Paranal Observatory, Chile, about 10~km from their final 
destinations at the CTAO-S site. The coordinates and altitudes are given in Table~\ref{tab:sites}.

\section{FRAM Purpose, Design and Data Acquisition} \label{sec:style}

FRAM (F/Photometric Robotic Atmospheric Monitor) is a series of robotic optical 
telescopes used for atmospheric monitoring at astroparticle 
physics observatories. FRAM is a passive instrument that does not produce any artificial light, it can 
operate within the field of view of other detectors without interference. At the 
Pierre Auger Observatory \citep{2015172} in Malarg\"{u}e, Argentina, FRAM has been operational 
since 2005 \citep{Aab_2021}, where it is used for cloud detection along cosmic-ray shower tracks 
and contributes to rapid atmospheric monitoring for fluorescence detector 
observations \citep{2015172}. At the future CTAO sites, FRAM has been deployed for site 
characterization and will monitor atmospheric transparency for 
Cherenkov light detection \citep{ebr2019prototypeoperationsatmosphericcalibration}.

The data presented in this paper are from the FRAMs at CTAO-N and at the intermediate Paranal location of CTAO-S0 and CTAO-S1. 
The CTAO FRAMs use a 135~mm f/2 telephoto lens (7~cm aperture diameter) 
coupled with a Moravian Instruments G4-16000 monochrome 
CCD camera featuring a $36\times36$~mm sensor with 16 megapixels, 
operated at a stabilized temperature of $-20$\degr C. The lens is 
a Zeiss or Samyang model depending on the site and period: the 
Zeiss lens was replaced by a Samyang lens at CTAO-N on 2021 
September 26 and at CTAO-S0 on 2022 March 24, while CTAO-S1 
retained the Zeiss lens throughout the period covered by this work. 
The camera is mounted on a robotic equatorial mount, allowing automated 
tracking and positioning across the sky, with exposure start times 
synchronized to UTC via NTP protocol to an accuracy better than 1 second. This 
combination provides a wide field of view of $15\times15$ degrees with a pixel 
scale of 14.1\arcsec/pixel, designed to cover the full field of view of the 
Imaging Atmospheric Cherenkov Telescopes (IACTs) with a significant margin. 
The cameras are equipped with photometric Johnson-Cousins $B$, $V$, and $R$ 
filters. The effective wavelengths of the $B$, $V$, and $R$ 
filters were determined by computing the flux-weighted mean wavelength, 
integrating the product of the filter transmission, CCD quantum efficiency, 
lens transmission, and the Vega spectrum over the filter bandpass. The 
resulting values were rounded to the nearest 10~nm, giving effective 
wavelengths of 440, 540, and 620~nm for the $B$, $V$, and $R$ filters 
respectively. At the Paranal Observatory in Chile, two identical FRAMs are 
installed 1.3 km apart from each other, while a single FRAM is 
located at the Observatorio del Roque de los Muchachos in La Palma, Spain.

\begin{table}[htbp]
\centering
\caption{Locations of the FRAMs whose data are used in this work.}
\setlength{\tabcolsep}{4pt}
\begin{tabular}{lccc}
\hline
Site & Latitude & Longitude & Altitude (m a.s.l.) \\
\hline
CTAO-N  & $28.7623^{\circ}$ N & $17.8900^{\circ}$ W & 2185 \\
CTAO-S0 & $24.6409^{\circ}$ S & $70.3870^{\circ}$ W & 2380 \\
CTAO-S1 & $24.6337^{\circ}$ S & $70.3770^{\circ}$ W & 2345 \\
\hline
\end{tabular}
\label{tab:sites}
\end{table}

Our method measures the integrated vertical aerosol optical depth (VAOD) at 
zenith. A single VAOD measurement is obtained from an altitude scan consisting 
of a sequence of 8 images taken at increasing altitudes. The 
scan begins near the horizon and progresses towards the zenith in steps of 10 
degrees in elevation. At each 
altitude, images are taken in all three filters ($B$, $V$, and $R$) 
at a fixed azimuth angle. In addition, $B$-filter scans are also taken at varying azimuths. This scan sequence was used to obtain the data analysed in this paper; it is not the method foreseen for final operation with CTAO. The airmass model assumes a stratified atmosphere, which is most probably valid at Paranal but not at CTAO-N \citep{Fruck2022, Gaug2022}.


\section{Transition from Tycho-2 to Gaia DR3}
\label{sec:transition}

\subsection{Extension to $V$ and $R$ Filters}

The method measures VAOD by comparing the instrumental magnitudes 
of stars observed at different altitudes to their catalogue 
magnitudes. The difference between the two is modelled as a function 
of airmass, from which the aerosol extinction is extracted by fitting 
the magnitude equation (described in detail in 
Section~\ref{sec:floats}). The original pipeline was developed and 
optimised for the $B$ filter using Tycho-2 as the reference 
catalogue \citep{Ebr2021}, with an aperture radius of 4~pixels, a 
neighbour rejection radius of 20~pixels, star selection constraints 
on magnitude range ($6.5 < B < 9.5$), colour index ($B-V < 0.85$) 
and FWHM ($\mathrm{FWHM} < 5$~pixels), and a fixed airmass 
averaging coefficient of 0.24. The Tycho-2 catalogue provides 
magnitudes in the $B_T$ and $V_T$ passbands, which require a 
transformation to the Johnson system and do not include an $R$ 
filter, making it unsuitable as a reference for $V$ and $R$ filter 
photometry. Gaia~DR3 provides synthetic photometry directly in the 
Johnson--Cousins $B$, $V$, and $R$ system, making it possible to 
extend the VAOD measurement to all three filters simultaneously 
without any passband transformation. This extension required several 
adjustments to the pipeline that are discussed in the following 
subsections.

\subsubsection{Field Crowding in V and R Filters}

\begin{figure}[htbp]
    \centering
    \includegraphics[width=\columnwidth]{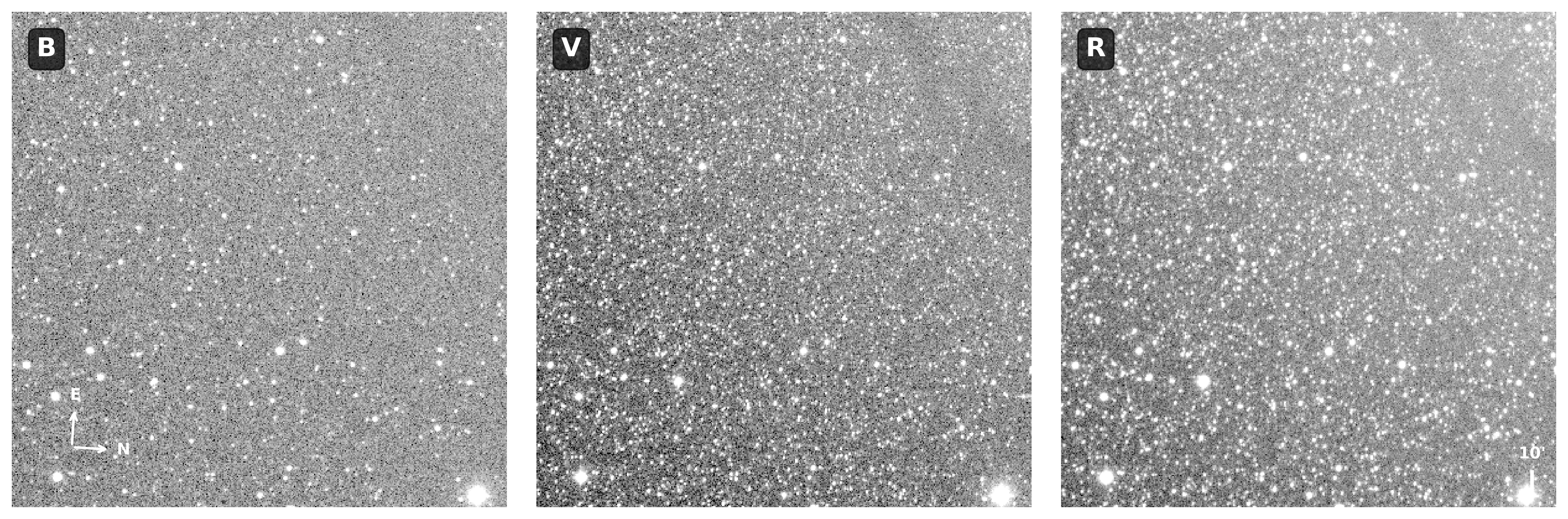}
    \caption{Comparison of detected sources in $B$, $V$, and $R$ filter images 
from the same scan shown in Figure~\ref{fig:extinction_plots}. All three 
images were taken consecutively at the same pointing (altitude $27^{\circ}$) 
with identical exposure times. The $B$-filter image (left) shows fewer 
detected stars due to stronger atmospheric extinction and lower system 
sensitivity at shorter wavelengths, while $V$ (center) and $R$ (right) filter 
images appear increasingly crowded.}
    \label{fig:BVR_field_comp}
\end{figure}

Rayleigh scattering scales as $\lambda^{-4}$, while aerosol scattering scales 
approximately as $\lambda^{-\alpha}$ where $\alpha$ is typically between 0 and 
2. Under clear-sky conditions, Rayleigh scattering dominates. Since the $B$-filter is centered at the shortest wavelength (440\,nm), it experiences the strongest atmospheric extinction and 
the poorest sensitivity of the lens--camera system. This means that for the same exposure time, fewer faint 
stars are detected in the $B$-filter compared to $V$ and $R$ filters . Consequently, images taken in $V$ and $R$ filters 
appear more crowded with detected sources (see Figure~\ref{fig:BVR_field_comp}).

Originally the pipeline rejected stars with a neighbouring source within 20 pixels 
to avoid blending effects in photometry. While this criterion worked well for 
the less crowded $B$-filter images, applying it to $V$ and $R$-filter data 
rejected too many stars, leaving insufficient measurements to reliably identify 
cloud-contaminated scans from the extinction profile. We therefore reduced the 
neighbour rejection distance to 10~pixels for all Gaia~DR3 processing across 
all three filters. To quantify this change, we processed the data from the 
period 2021 September 26 to 2021 October 17 (the yellow highlighted region in 
Figure~\ref{fig:vaod}) at CTAO-S0 in all three filters with both distances, 
all other pipeline configuration parameters being identical. For the 59 scans 
common to all filters and both configurations, reducing the distance from 20 
to 10~pixels increases the mean number of stars used per scan by a factor of 
1.2 in the $B$ filter (from about 6,800 to 8,500), 4.3 in the $V$ filter 
(from about 2,100 to 9,200), and 5.7 in the $R$ filter (from about 2,700 to 
15,400).

In addition to the neighbour rejection distance, two further parameters govern the detection and matching of stars: a detected source is matched to a catalogue star only if the closest counterpart lies within 8~pixels, ensuring reliable cross-identification without spurious matches; and the sky background for each star is estimated from an annular region spanning 10 to 20~pixels from the star centre, outside the photometric aperture.

\begin{figure}[htbp]
    \centering
    \begin{subfigure}{0.32\textwidth}
        \centering
        \includegraphics[width=\textwidth]{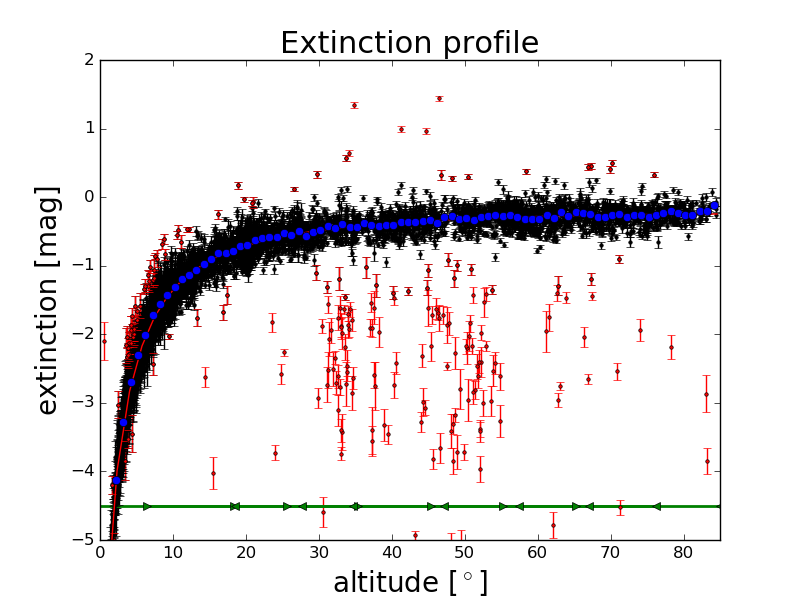}
        \caption{B filter}
        \label{fig:extinction_B}
    \end{subfigure}
    \hfill
    \begin{subfigure}{0.32\textwidth}
        \centering
        \includegraphics[width=\textwidth]{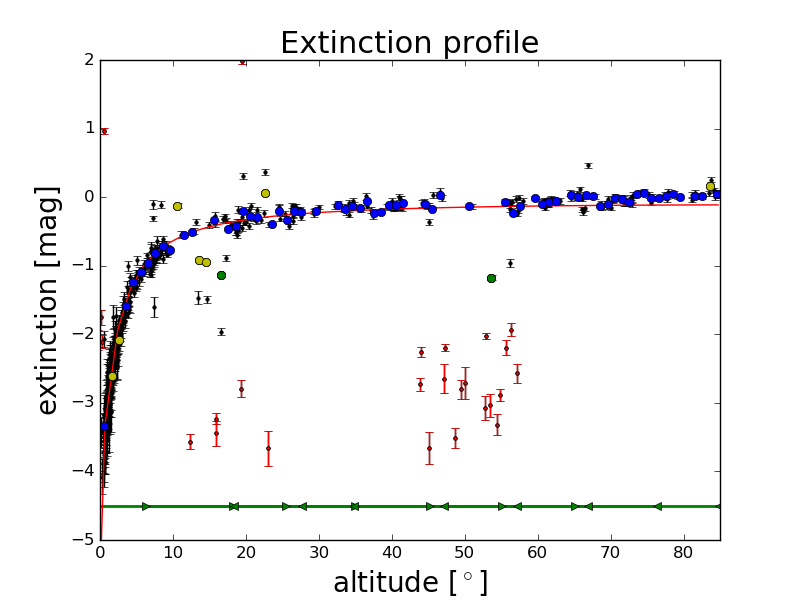}
        \caption{V filter}
        \label{fig:extinction_V}
    \end{subfigure}
    \hfill
    \begin{subfigure}{0.32\textwidth}
        \centering
        \includegraphics[width=\textwidth]{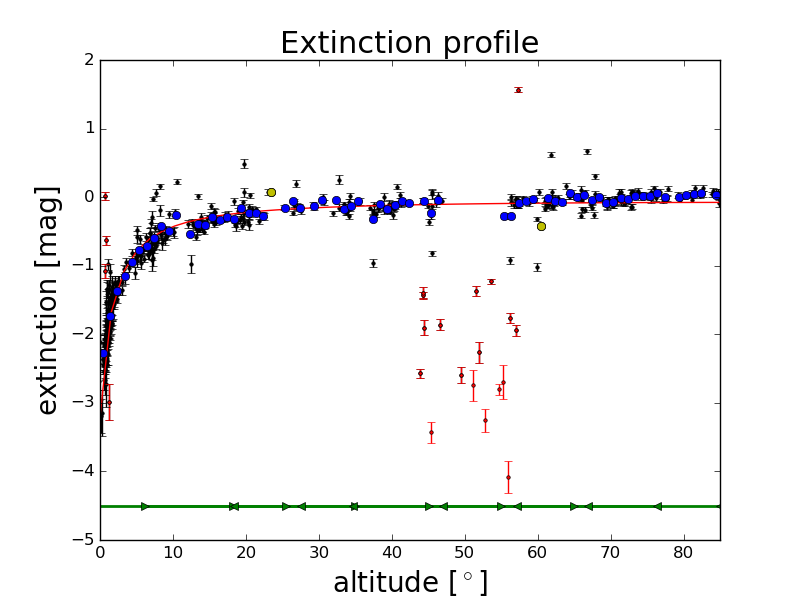}
        \caption{R filter}
        \label{fig:extinction_R}
    \end{subfigure}
    \caption{Extinction profiles from preliminary analysis for the CTAO-S0 site, 
    processed with the 20-pixel neighbour rejection distance and used to identify 
    and discard scans affected by clouds. The blue points 
    follow the fitted extinction model, while red points indicate stars marked 
    for rejection as outliers. The green arrows at the bottom show the altitude 
    coverage of individual images. Point colours reflect the mean RMS deviation 
    of stars in each altitude bin from the preliminary fit.}
    \label{fig:extinction_plots}
\end{figure}

Figure~\ref{fig:extinction_plots} shows extinction profiles from the same 
scan presented in Figure~\ref{fig:BVR_field_comp}, processed with the 
original 20-pixel neighbour rejection distance and used in the preliminary 
analysis to identify scans affected by clouds. Since the $V$ and $R$ fields 
are more crowded, a larger fraction of their stars have a neighbour within 
20 pixels and are rejected, leaving the $V$ and $R$ extinction profiles 
(Figures~\ref{fig:extinction_V} and~\ref{fig:extinction_R}) sparser than 
the $B$ profile (Figure~\ref{fig:extinction_B}).

\subsubsection{Star Selection Constraints}
\label{sec:starselection}

For the Tycho-2 pipeline, stars were required to satisfy the following 
constraints: a magnitude range ($6.5 < B < 9.5$), a full width at half 
maximum of the stellar profile ($\mathrm{FWHM} < 5$~pixels), a colour 
index ($B-V < 0.85$), an airmass range ($1 < \mathrm{airmass} < 8$), 
and a requirement that stars fall within the region where the PSF retains 
sufficient symmetry and is not significantly distorted by the lens optics 
($r < 1840$~pixels from the image centre). The magnitude range was imposed 
to avoid saturated stars at the bright end and to exclude faint stars where 
the photometric precision of Tycho-2 degrades significantly \citep{Hog2000}. 
The colour index limit was required because the colour correction polynomial 
becomes unreliable for red stars in the Tycho-2 system \citep{Ebr2017}. The 
FWHM cut rejected stars with poor PSF profiles that could contaminate 
aperture photometry.

For Gaia~DR3, only the airmass range and the detector area constraint are 
retained. The strict magnitude range is not applied since the Gaia~DR3 
synthetic photometry remains reliable to considerably fainter magnitudes 
and the synthetic photometry shows little dependence on colour index across 
the full stellar colour range, making the colour index cut unnecessary. We 
remove the FWHM cut since with the substantially larger number of stars 
available per scan with Gaia~DR3 (Figure~\ref{fig:mag_histogram}), the 
contribution of any individual poorly-resolved star to the fit is negligible.

\begin{figure*}[htbp]
    \centering
    \includegraphics[width=\textwidth]{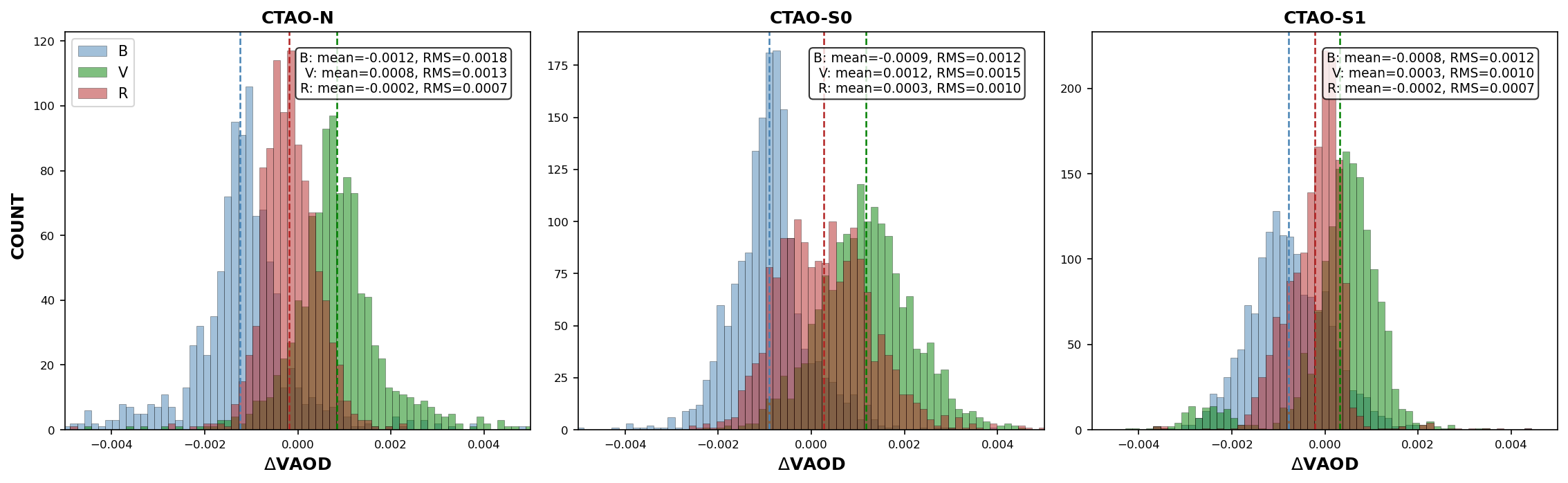}
    \caption{Difference in VAOD between the standard Gaia~DR3 star selection 
(airmass and detector area constraints only) and a more restrictive 
selection additionally requiring $\mathrm{FWHM} < 5$~pixels and 
$B-V < 0.85$, for all three CTAO sites and all three filters ($B$, $V$, 
$R$).}
    \label{fig:vaod_CI_fwhm}
\end{figure*}

To verify that removing these constraints does not introduce a significant 
bias in the derived VAOD, we compared results with and without the FWHM 
and colour index cuts applied to the Gaia~DR3 data for all three filters 
over the full data period 2021 January 1 to 2022 December 31. The mean difference and the RMS are both below $0.002$ at all sites and in all three filters (Figure~\ref{fig:vaod_CI_fwhm}), with the $B$ filter showing 
the largest spread and the $R$ filter the smallest, confirming that these 
constraints have a negligible effect on the final VAOD values and 
justifying their removal from the Gaia~DR3 pipeline.

\subsubsection{Colour Correction and Residual Colour Dependence}
\label{sec:colorcorr}

For Tycho-2, the $B_T$ and $V_T$ passbands do not exactly correspond 
to the Johnson $B$ and $V$ filters, requiring a colour-dependent 
transformation to convert to the Johnson system \citep{Hog2000}. This 
transformation introduces a significant residual dependence of the 
model-minus-instrumental magnitude on stellar colour index. For Gaia~DR3, 
the synthetic photometry is computed directly in the 
Johnson--Cousins system, so no passband transformation is needed. 
However, even with accurate catalogue photometry, a small residual 
dependence on stellar colour index remains because the overall 
effective response of the FRAM system --- determined by the combined 
transmission of the lens, filter, and CCD --- is not exactly the 
ideal Johnson--Cousins passbands assumed in the synthetic photometry. In 
both cases this residual is accounted for by the colour correction 
polynomial term $f$ in the magnitude equation 
(Equation~\ref{eq:mag}).

\begin{figure}[htbp]
    \centering
    \includegraphics[width=0.65\columnwidth]{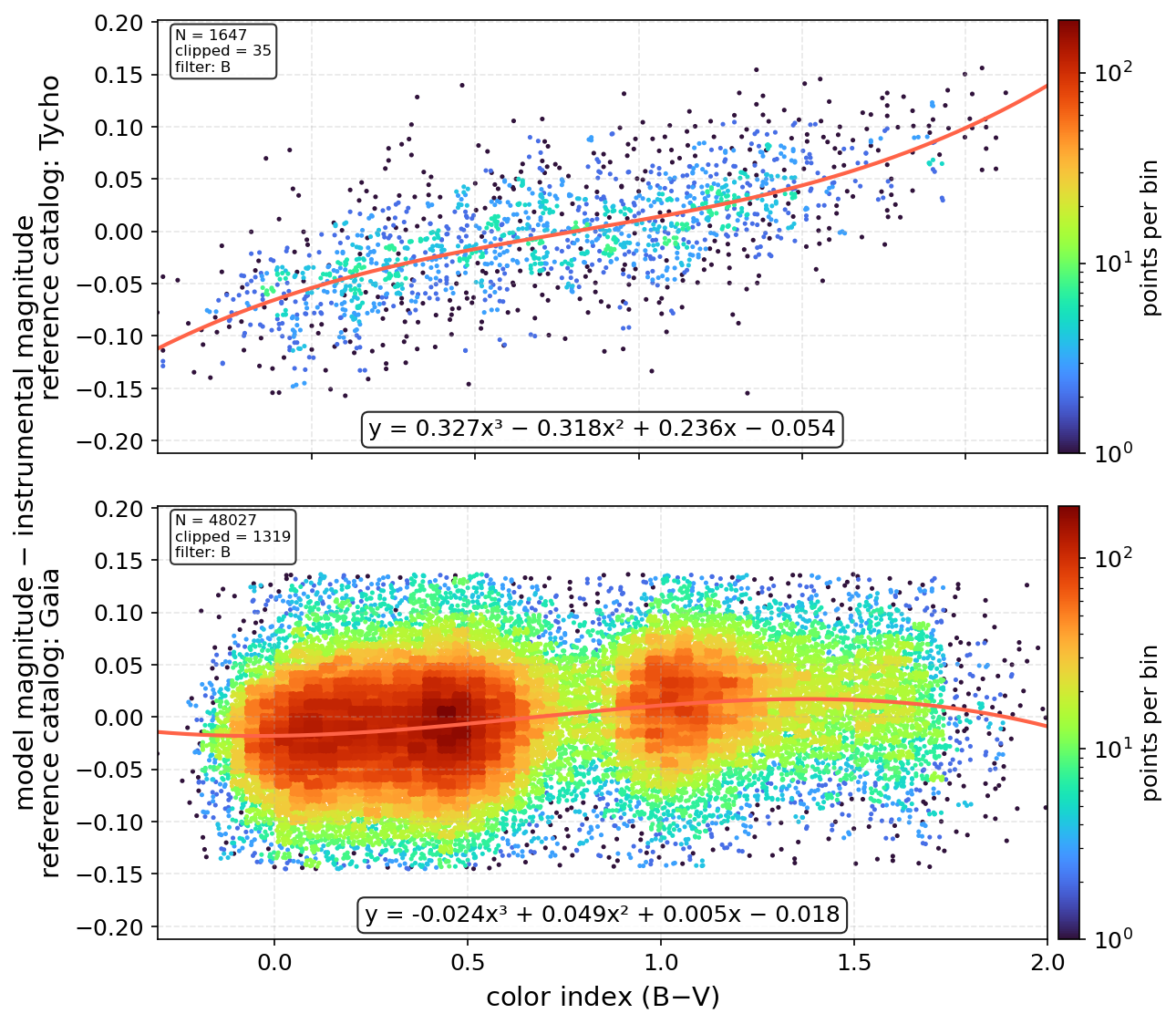}
    \caption{Residual between model and instrumental magnitude as a function
    of colour index ($B-V$) for the $B$ filter at CTAO-S0, using Tycho-2
    (top) and Gaia~DR3 (bottom). Points are coloured by local density
    (number of points per bin); the red curve is a cubic polynomial fit to
    the $3\sigma$-clipped data. The fitted coefficients are shown in each
    panel.}
    \label{fig:colorcorr_B}
\end{figure}

\begin{figure}[htbp]
    \centering
    \includegraphics[width=0.65\columnwidth]{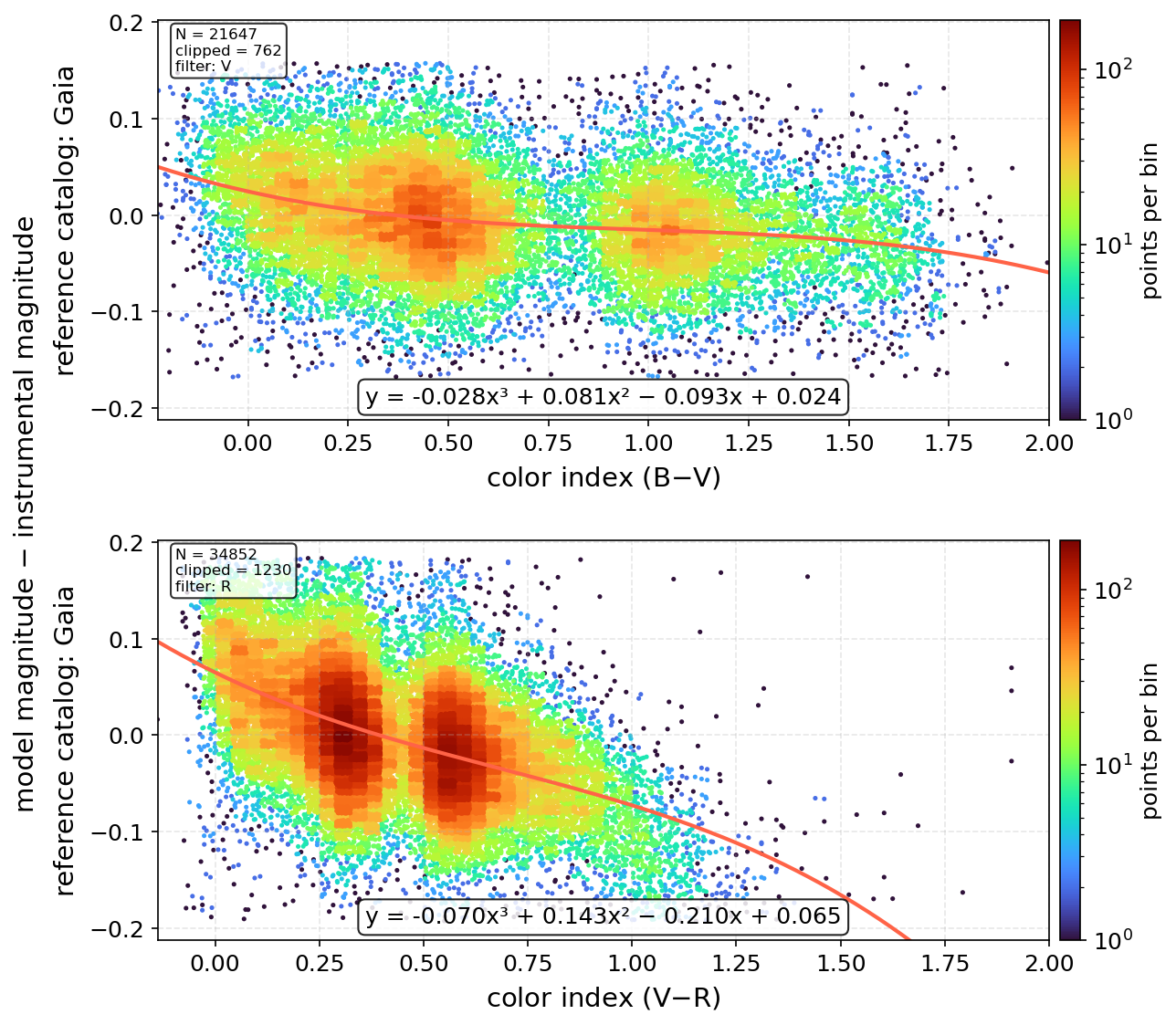}
    \caption{Same as Figure~\ref{fig:colorcorr_B} but for the $V$ filter
    (top, colour index $B-V$) and $R$ filter (bottom, colour index $V-R$)
    using Gaia~DR3.}
    \label{fig:colorcorr_VR}
\end{figure}

Figures~\ref{fig:colorcorr_B} and~\ref{fig:colorcorr_VR} show the 
residual between the model and instrumental magnitude as a function of 
colour index for all three filters, derived from five scans at CTAO-S0 
from early 2021, which form part of the full dataset presented in 
Figure~\ref{fig:vaod}. For the $B$ filter, the residual is shown for both 
Tycho-2 and Gaia~DR3, while for the $V$ and $R$ filters only Gaia~DR3 is shown since Tycho-2 
provides no $R$ photometry and the $V$ filter result is similar to $B$. 
For Tycho-2, the residual shows a strong and systematic colour dependence 
across the limited range $B-V \in [-0.2, 0.9]$. When Gaia~DR3 is used 
instead, the colour dependence in the $B$ filter is dramatically reduced 
and the cubic fit is nearly flat across a much wider range 
$B-V \in [-0.5, 2.0]$. The $V$ filter shows a similarly small residual 
colour dependence.


\subsubsection{Optimum Aperture Selection}
\label{sec:aperture}

The photometric pipeline extracts stellar flux using circular aperture
photometry, and the choice of aperture radius affects the accuracy of
the derived VAOD values. The fitting procedure determines a single
zero-point $Z_i$ and a single aerosol extinction coefficient $k$ for
each scan. A uniform flux loss due to the aperture choice would be
accounted for by $Z_i$ and would not affect the derived VAOD. However,
the flux loss is not uniform across stars --- it varies depending on
the PSF shape and position in the field of view. Since $Z_i$ can only
account for a constant offset and $k$ only describes airmass-dependent
extinction, this non-uniform flux loss cannot be accounted for by any
fitted parameter and remains as a residual after the fit. The optimum
aperture is therefore defined as the one that minimises the mean RMS
of the residuals between the measured and model-predicted magnitudes.

\begin{figure*}[htbp]
    \centering
    \includegraphics[width=0.75\textwidth]{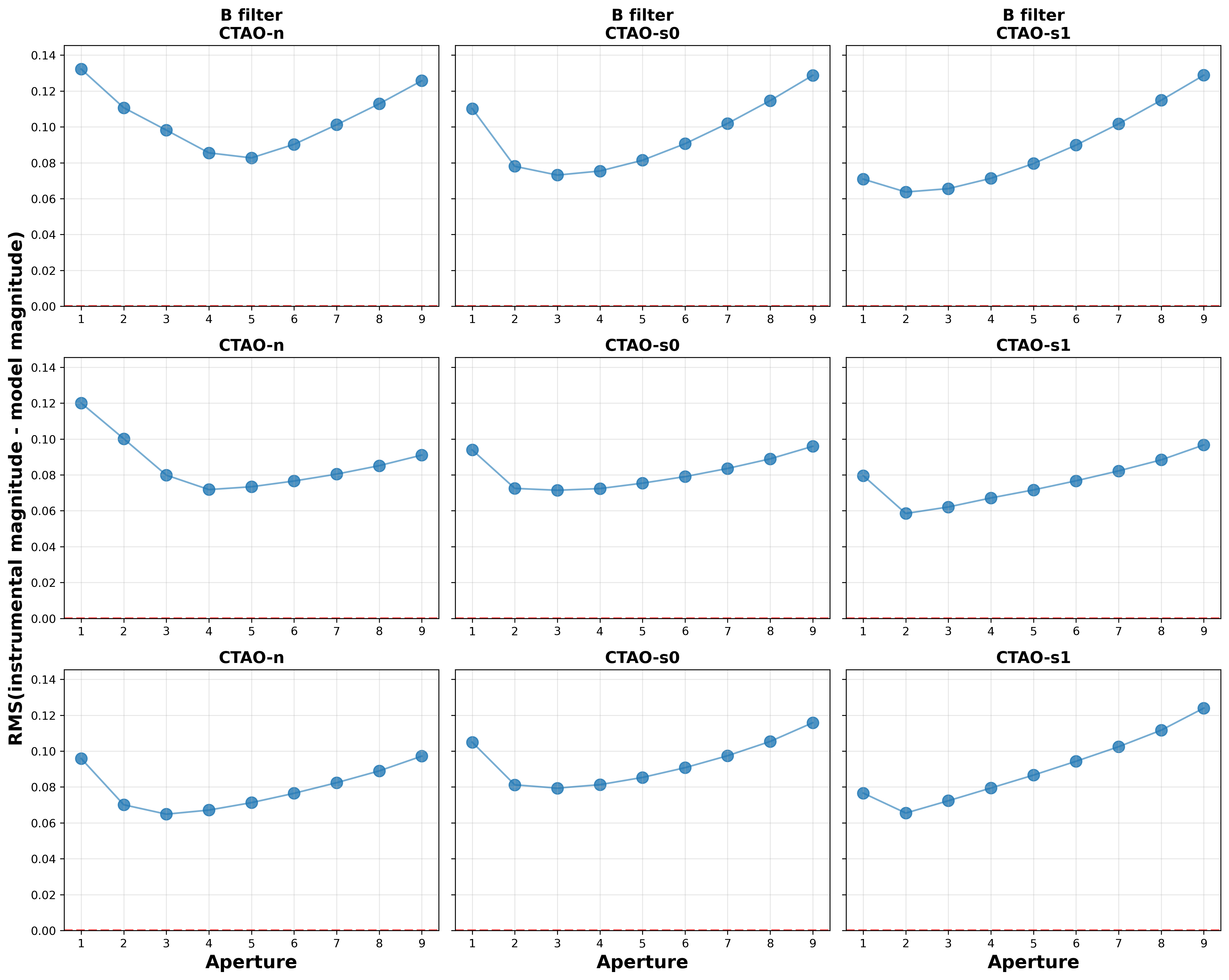}
    \caption{Mean RMS of residuals between the measured and
    model-predicted magnitudes as a function of aperture radius
    (1--9 pixels) for the three sites (rows) and three filters
    (columns), computed from 16 scans at CTAO-N, 26 at CTAO-S0,
    and 28 at CTAO-S1 from the period 2022 February 01 to 2022
    April 01.}
    \label{fig:aperture}
\end{figure*}

To determine this aperture, we used data from the period 2022
February 01 to 2022 April 01, corresponding to the purple
highlighted region in Figure~\ref{fig:vaod}. Each observation
consists of a sequence of images taken at a fixed azimuth, starting
at an altitude of $7^{\circ}$ and increasing in $10^{\circ}$ steps
to a maximum altitude of $77^{\circ}$. All images in a sequence
share a common scan identifier across all three filters. For this
analysis, we restricted the sample to scans that were successfully
processed in all three filters ($B$, $V$, $R$) simultaneously,
yielding 16 scans for CTAO-N, 26 for CTAO-S0, and 28 for CTAO-S1.

The mean RMS was computed for aperture radii ranging from 1 to
9 pixels, and the results are shown in Figure~\ref{fig:aperture}.
The minimum is stable across filters at CTAO-S0 (aperture 3) and
CTAO-S1 (aperture 2), but at CTAO-N it shifts with wavelength,
from aperture 5 for the $B$ filter to 4 for the $V$ filter and 3
for the $R$ filter. Since no single aperture minimises the
residual RMS simultaneously across all sites and filters, and
since repeating the analysis on different data subsets shifts the
minima further, we adopted 3 pixels at CTAO-N and 2 pixels at
CTAO-S0 and CTAO-S1, the same for all three filters, as
reasonable, workable values applicable to the full multi-year
dataset, rather than a strictly filter- and epoch-dependent optimum. Section~\ref{sec:config_aper5} shows that
fixing the aperture at 5~pixels instead changes the derived VAOD
by only a small amount at every site, confirming that this choice
has little practical effect on the results.

A common aperture set is also motivated by the evolution of the instrumental systems over the multi-year dataset, so that no single aperture is strictly optimal at all times, and it ensures that comparisons between filters are not affected by aperture-related systematics.

\subsubsection{Aperture Dependence of VAOD}
\label{sec:aperture_dependence}

\begin{figure*}[htbp]
    \centering
    \includegraphics[width=\textwidth]{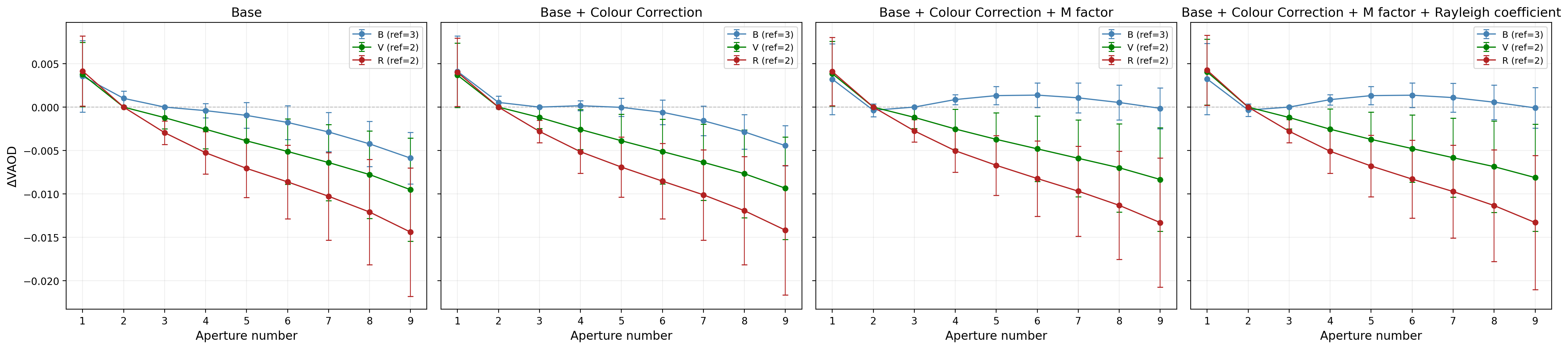}
    \caption{$\Delta\mathrm{VAOD} = \mathrm{VAOD}(\mathrm{aperture}) - 
    \mathrm{VAOD}(\mathrm{ref})$ as a function of aperture number 
    (1--9~pixels) for the $B$, $V$, and $R$ filters (overlaid, with 
    error bars showing the standard deviation), across the four configurations of increasing complexity listed in the text (left to right). The reference aperture 
    for each filter is the optimum aperture determined in 
    Section~\ref{sec:aperture} (3, 2, and 2~pixels for $B$, $V$, and 
    $R$ respectively). Data are from CTAO-S0 over the period 2021 
    September 26 to 2021 October 17, restricted to the 48 observations 
    common to all apertures and configurations.}
    \label{fig:vaod_aperture}
\end{figure*}

\begin{table}[htbp]
    \centering
    \caption{Filter-specific Rayleigh extinction coefficients used for 
    airmass averaging at each CTAO site, derived from the MODTRAN 
    molecular atmosphere calculation at each site altitude.}
    \begin{tabular}{lccc}
        \hline
        Site & $B$ & $V$ & $R$ \\
        \hline
        CTAO-N  & 0.221 & 0.093 & 0.063 \\
        CTAO-S0 & 0.215 & 0.091 & 0.061 \\
        CTAO-S1 & 0.216 & 0.091 & 0.061 \\
        \hline
    \end{tabular}
    \label{tab:rayleigh_coeffs}
\end{table}

\begin{table*}[htbp]
\centering
\caption{Summary of the final Gaia~DR3 pipeline configuration adopted 
for all VAOD measurements presented in this work, compared with the 
original Tycho-2 $B$-filter configuration.}
\setlength{\tabcolsep}{3pt}
\begin{tabular}{lll}
\hline
Parameter & Gaia~DR3 & Tycho-2 \citep{Ebr2021} \\
\hline
Reference catalogue & Gaia~DR3 synthetic photometry & Tycho-2 \\
Photometric aperture radius & 3~pixels (CTAO-N), 2~pixels (CTAO-S0, CTAO-S1), all filters & 4~pixels \\
Neighbour rejection radius & 10~pixels & 20~pixels \\
Source matching radius & 8~pixels & 8~pixels \\
Sky background annulus & 10--20~pixels from star centre & 10--20~pixels from star centre \\
Colour correction & Cubic polynomial ($3\sigma$-clipped) & Cubic polynomial ($3\sigma$-clipped) \\
Non-linearity correction & $M$ factor applied to full & $M$ factor applied to \\
                                           & magnitude equation (Equation~\ref{eq:mag}) & catalogue magnitude only \\
Rayleigh coefficients & Filter- and site-specific (Table~\ref{tab:rayleigh_coeffs}) & Fixed value of 0.24 \\
Star selection & Airmass ($1 < \mathrm{airmass} < 8$) and & Magnitude ($6.5 < B < 9.5$), FWHM ($<5$~pixels), \\
                                 & detector area constraint only & colour index ($B-V < 0.85$), airmass, detector area \\
Aerosol airmass model & Tropospheric ($A_A$; Section~\ref{sec:aerosol_ext}) & Tropospheric ($A_A$) and stratospheric ($A_O$) \\
\hline
\end{tabular}
\label{tab:final_config}
\end{table*}

We assess how sensitive the derived VAOD is to deviations from the 
optimum aperture (3, 2, and 2~pixels for $B$, $V$, and $R$ respectively), 
and whether this sensitivity is reduced by the successive pipeline configuration 
corrections. We processed observations taken between 2021 September 26 
and 2021 October 17 at CTAO-S0 (yellow highlighted region in 
Figure~\ref{fig:vaod}) across four configurations of increasing 
complexity. The base configuration uses the following settings: the sky 
background is estimated from an annular region spanning 10 to 20~pixels 
from the star centre; a detected source is matched to a catalogue star 
only if the closest counterpart lies within 8~pixels; any source with a 
second catalogue star within 10~pixels is rejected to avoid flux 
contamination from blended neighbours; and the airmass averaging 
coefficient is set to 0.24, a value originally adopted from the 
$B$-filter Tycho-2-based processing.

Two of the four configurations involve parameters that require brief 
explanation. The airmass averaging coefficient is an approximate 
atmospheric optical depth used to compute the effective airmass over 
the finite exposure duration: rather than averaging the airmass directly, 
the atmospheric transmission is averaged across the exposure and 
converted back to an effective airmass, accounting for the exponential 
nature of atmospheric extinction. The magnitude scaling factor $M$ 
corrects for any residual non-linearity between the instrumental and 
catalogue magnitude scales not removed during image calibration; it is 
fitted as a single global parameter and is expected to be close to 
unity. Because $M$ relates the instrumental magnitude scale to the true scale, it multiplies the entire model prediction, including the zeropoint, extinction, and colour terms; applying it only to the catalogue magnitude leaves the other terms unscaled and introduces an aperture-dependent bias in the derived VAOD. The four configurations are:

\begin{enumerate}
    \item \textbf{Base}: the base configuration.
    \item \textbf{Base + colour correction}: a polynomial colour 
    correction is additionally applied to the instrumental magnitudes.
    \item \textbf{Base + colour correction + $M$ factor}: the $M$ 
    factor is additionally applied to all terms in the magnitude 
    equation (Equation~\ref{eq:mag}) rather than only to the catalogue 
    magnitude.
    \item \textbf{Base + colour correction + $M$ factor + Rayleigh 
    coefficient}: the airmass averaging coefficient is additionally 
    replaced with the filter-specific approximate Rayleigh extinction 
    coefficient (0.215, 0.091, and 0.061 for $B$, $V$, and $R$ 
    respectively at CTAO-S0; see Table~\ref{tab:rayleigh_coeffs}), 
    rather than the fixed value of 0.24.
\end{enumerate}

For each configuration and filter, we compute 
$\Delta\mathrm{VAOD} = \mathrm{VAOD}(\mathrm{aperture}) - 
\mathrm{VAOD}(\mathrm{ref})$ for aperture radii 1--9~pixels, where the 
reference is the optimum aperture for that filter, using only 
observations common to all apertures and all four configurations. 
Figure~\ref{fig:vaod_aperture} shows the resulting $\Delta\mathrm{VAOD}$ 
as a function of aperture number for all three filters overlaid in each 
panel. In the base configuration (first panel), $\Delta\mathrm{VAOD}$ 
decreases steeply and monotonically with aperture for all three filters, 
indicating that small apertures are more susceptible to background 
contamination and PSF-related flux losses. The addition of the $M$-factor 
correction (third panel) has the largest individual impact on the $B$ 
filter, flattening its $\Delta\mathrm{VAOD}$ profile and substantially 
reducing its maximum RMS (from 0.0066 in the base configuration to 0.0052 
in the final configuration). The $V$ and $R$ filters, in contrast, show 
little improvement from the successive corrections: their 
$\Delta\mathrm{VAOD}$ profiles remain steep and monotonic in all four 
configurations, with maximum RMS only marginally reduced (from 0.0112 to 
0.0102 for $V$, and from 0.0162 to 0.0154 for $R$). The corrections therefore reduce the aperture sensitivity of the $B$ filter substantially but leave $V$ and $R$ considerably more sensitive even in the final configuration.

\subsubsection{Final Pipeline Configuration}
\label{sec:final_config}

The final configuration adopted for all 
subsequent analysis in this paper is summarised in 
Table~\ref{tab:final_config}.

\subsection{VAOD Consistency Between Tycho-2 and Gaia DR3}
\label{sec:tyco_gaia_comparison}

We compare the VAOD values derived using Gaia~DR3 with those obtained using the original Tycho-2 catalogue and its corresponding configuration. This comparison is performed for the $B$ filter at all three CTAO sites, since Tycho-2 provides no $R$ photometry and the method for $V$ filter using Tycho-2 was never developed.

To assess the consistency of VAOD values 
derived from the two catalogues, we compared Tycho-2 and Gaia~DR3 VAOD 
measurements for matched scans at all three CTAO sites over the period 
2021 January 1 to 2022 December 31.

\begin{figure*}[htbp]
    \centering
    \includegraphics[width=\textwidth]{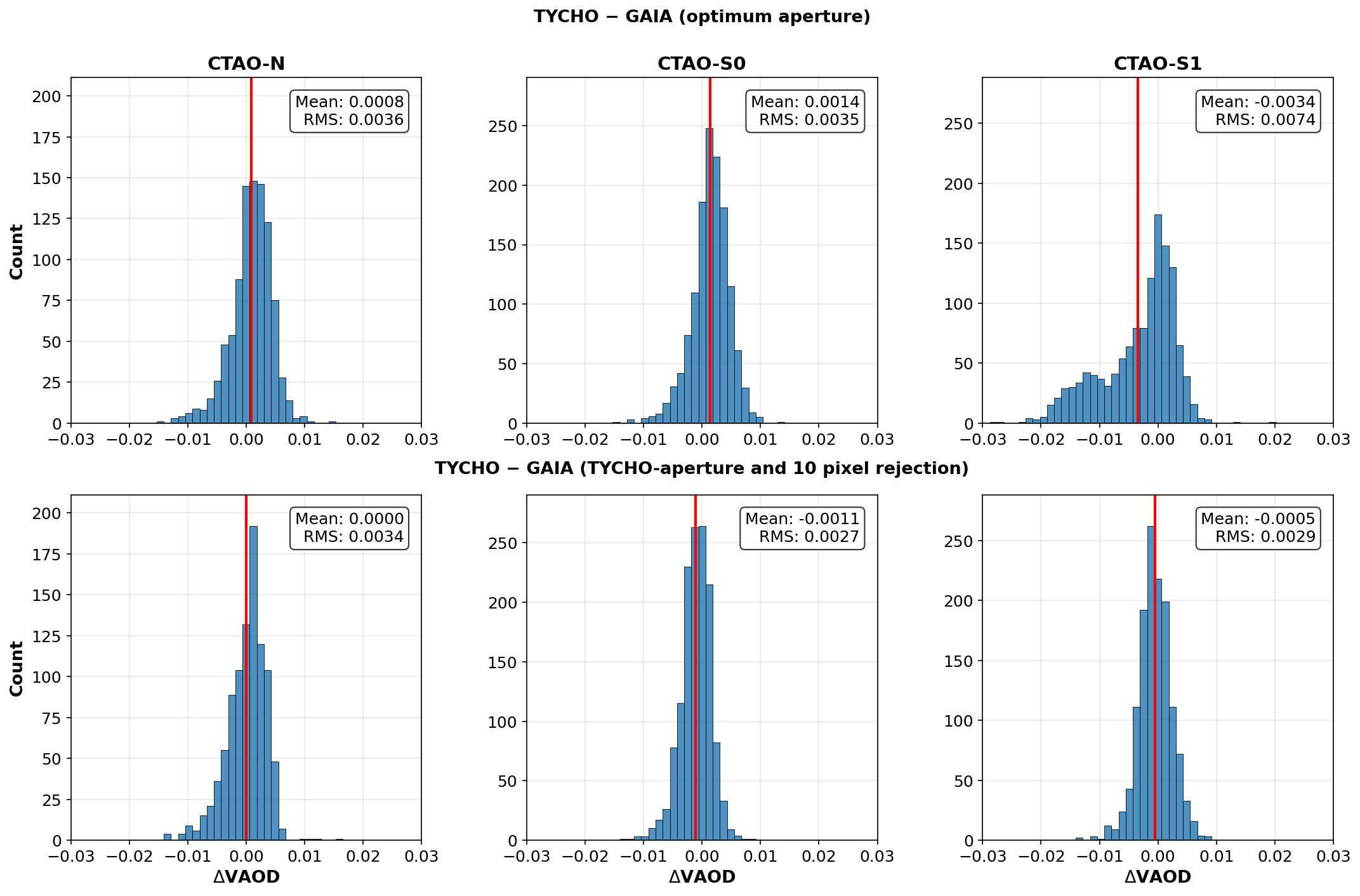}
    \caption{Distributions of the difference in VAOD between Tycho-2 and 
    Gaia~DR3 for matched scans at CTAO-N, CTAO-S0, and CTAO-S1 (columns) 
    over the period 2021 January 1 to 2022 December 31, for two Gaia~DR3 
    configurations: the optimum aperture with the standard Gaia~DR3 neighbour 
    rejection criterion (top row), and the Tycho-2 aperture with a tighter 
    neighbour rejection criterion (bottom row). The secondary population 
    visible at CTAO-S1 in the top row defines the outlier scans discussed in 
    the text.}
    \label{fig:vaod_tycho_gaia}
\end{figure*}

As shown in Figure~\ref{fig:vaod_tycho_gaia}, the agreement between 
the two catalogues is good at CTAO-N and CTAO-S0. At CTAO-S1, however, 
the Gaia~DR3 configuration 
with the optimum aperture of 2~pixels produces a broader and asymmetric 
VAOD difference distribution ($\mathrm{RMS} = 0.0074$, mean $= -0.0034$), 
with a secondary population of scans where $|\Delta\mathrm{VAOD}| > 0.01$, 
absent in the Tycho-2 comparison. We refer to these as outlier scans. 
To investigate the origin of this secondary population, we tested two 
additional Gaia~DR3 configurations: one using the Tycho-2 aperture of 
4~pixels \citep{Ebr2021} instead of the optimum 2~pixels, and one using 
the Tycho-2 aperture combined with a tighter neighbour rejection 
criterion. The latter showed no significant difference with respect to 
the former, confirming that the neighbour rejection criterion does not 
contribute to the secondary population. When the Tycho-2 aperture of 
4~pixels is used, the secondary population disappears 
($\mathrm{RMS} \approx 0.003$, mean $\approx -0.001$), indicating that 
the aperture size is the primary cause of the discrepancy at CTAO-S1.

Further analysis of the outlier scans reveals that they are not uniformly 
distributed in time. When examined as a time series, the outlier scans are found to be concentrated in the period 
before the end of 2021, after which they disappear abruptly. This 
temporal boundary coincides with a measurable improvement in the median 
FWHM of stellar profiles in the images, most likely due to a change in 
the focusing routine at the site, though this could not be confirmed 
with certainty. Since the two catalogue configurations use different 
aperture sizes --- 2~pixels for Gaia~DR3 and 4~pixels for Tycho-2 --- 
a change in PSF quality affects the flux measured in each aperture 
differently, producing a systematic difference in the derived VAOD 
between the two processing chains. The outlier behaviour is therefore 
attributed to this PSF change at CTAO-S1 around the end of 2021 rather 
than to a catalogue-related systematic effect.

\subsection{Catalogue Depth and Star Statistics}
\label{sec:statstistics}

The number of stars available for VAOD fitting directly affects the reliability 
of the photometric solution --- more stars across a wider range of altitudes 
and colours reduce the statistical uncertainty in the fitted zero-point $Z_i$ 
and aerosol extinction $k$. To quantify the improvement from transitioning to 
Gaia~DR3, we compared the star counts from both catalogues using five matched 
scans at CTAO-S1.

\begin{figure}[htbp]
    \centering
    \resizebox{0.7\hsize}{!}
    {\includegraphics{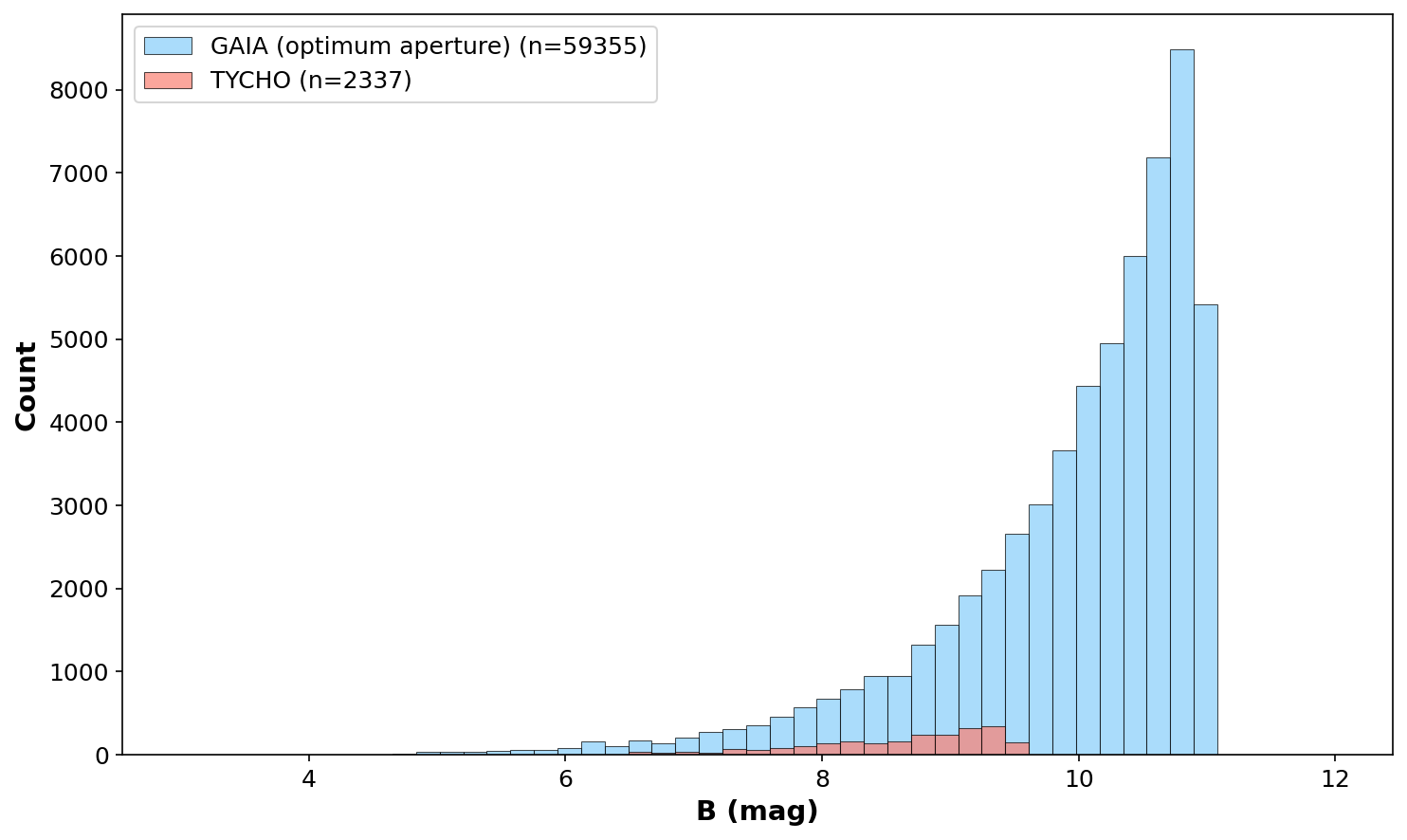}}
    \caption{Magnitude distribution of stars used in VAOD measurements for 
    five matched scans from CTAO-S1, comparing Gaia~DR3 (optimum aperture) 
    and Tycho-2. Gaia~DR3 includes $\sim$59,000 stars and Tycho-2 $\sim$2,300.}
    \label{fig:mag_histogram}
\end{figure}

Gaia~DR3 processing detected on average $\sim$11,870 stars per scan, while 
Tycho-2 processing detected only $\sim$467 stars per scan --- a ratio of 
approximately 25:1. The magnitude distributions of stars used in the VAOD 
measurements are shown in Figure~\ref{fig:mag_histogram}. The dominant factor driving this difference is the magnitude limit imposed on each catalogue: $6.5 < B < 9.5$ for Tycho-2, beyond which its photometry is not reliable, and $B < 11$ for Gaia~DR3. A secondary contribution comes from the neighbour rejection threshold, which was reduced from 20~pixels for Tycho-2 to 10~pixels for Gaia~DR3, retaining more stars in crowded fields. 
The significantly larger number of stars available with Gaia~DR3 improves the 
statistical precision of the fitted parameters and enables reliable VAOD 
measurements in the $V$ and $R$ filters.

\section{Photometric Method and Atmospheric Model} \label{sec:floats}

The method exploits the fact that Earth's atmosphere attenuates starlight as it 
passes through it. By comparing the measured instrumental magnitude of different 
stars at different altitudes above the horizon to their known catalogue magnitude, 
we can determine the amount of light lost to atmospheric extinction. We model this 
extinction as the sum of molecular scattering and absorption together with 
aerosol scattering \citep{Ebr2021}, each with a different wavelength and airmass dependence.

\subsection{Instrumental and Catalogue Magnitude Relation} 
\label{subsec:tables}

The measured instrumental magnitude of a star is related to its catalogue 
magnitude by:
\begin{equation}
\label{eq:mag}
m_{\text{inst}} = M \bigl[ m_{\text{cat}} + Z_i + g_M(A_M, \text{color-index}) + g_A(A_A, k_i, \text{color-index}) + f(\text{color-index}, \text{position}) \bigr]
\end{equation}
where $m_{\text{inst}}$ is the instrumental magnitude measured from the CCD 
image and $m_{\text{cat}}$ is the catalogue magnitude from Gaia~DR3 synthetic 
photometry in the Johnson--Cousins $B$, $V$, and $R$ filters. The term $f$ 
accounts for the colour-index- and position-dependent response of the instrument 
across the wide field of view. The term $g_M$ represents the atmospheric 
extinction from molecular scattering and absorption, computed from pre-calculated 
models as a function of airmass $A_M$ and stellar colour index. The aerosol 
 extinction term $g_A$ requires an assumed airmass function for the
aerosols, which in principle can be split into tropospheric and stratospheric
contributions. Since FRAM measures only the integrated optical depth and provides
no reliable way to separate the two, and in the absence of a stable stratospheric
contribution, we adopt the tropospheric airmass $A_A$ for the baseline and treat
the unknown vertical profile as a systematic uncertainty, evaluated in
Section~\ref{sec:configurations}. It depends on the aerosol optical depth 
$k$, the aerosol airmass $A_A$, and the colour index. The zero-point $Z_i$ is 
an additive constant fitted independently for each scan, absorbing any 
scan-to-scan variations in the absolute photometric scale. The non-linearity 
correction $M$ accounts for any brightness-dependent systematic in the detector 
response. The free parameters $Z_i$ and $k$ are determined for each scan by 
minimising the weighted sum of squared residuals between the measured 
and modelled magnitudes across all stars and altitudes in the scan.
To reject outliers, the fit is performed iteratively: after each iteration, 
stars with residuals exceeding $3\sigma$ are removed and the fit is repeated.

\subsection{Molecular Extinction}

We used MODTRAN \citep{Berk2014MODTRAN6} with atmospheric profiles from the 
ECMWF ERA5 dataset \citep{ECMWF_ERA5_2023} for June 20, 2018 (the effect of 
the time variation of the profile is assessed in Section~\ref{sec:mol_sensitivity}), 
to calculate the wavelength-dependent transmittance of the atmosphere. The 
following species were included in the calculation: the twelve default molecular 
species --- water vapor (H$_2$O), ozone (O$_3$), molecular oxygen (O$_2$), 
carbon dioxide (CO$_2$), methane (CH$_4$), nitrous oxide (N$_2$O), carbon 
monoxide (CO), nitric oxide (NO), nitrogen dioxide (NO$_2$), sulfur dioxide 
(SO$_2$), ammonia (NH$_3$), and nitric acid (HNO$_3$) --- along with peroxynitric 
acid (HNO$_4$), dinitrogen pentoxide (N$_2$O$_5$), and chlorine nitrate 
(ClONO$_2$). These species reduce the transmittance of the incoming light 
through line absorption at specific wavelengths. Of these molecular species, 
only ozone (O$_3$) and nitrogen dioxide (NO$_2$) contribute to absorption 
in the B and V filter wavelength ranges \citep{Voutsinas2026}. The calculation 
also accounted for continuum transmittance from water vapor (H$_2$O CONT) and 
nitrogen (N$_2$ CONT), as well as collision-induced absorption from H$_2$--H$_2$, 
H$_2$--He, H$_2$--CH$_4$, and CH$_4$--CH$_4$ pairs. Scattering transmittance 
due to molecular Rayleigh scattering by N$_2$ and O$_2$, which redirects photons 
out of the line of sight, was also incorporated. Various chlorofluorocarbons 
were included in the calculation but have negligible 
absorption in the 200--950~nm range. Aerosol and cloud transmittance were set 
to unity throughout, indicating clear-sky conditions. The total combined 
transmittance was calculated as the product of all individual line, continuum, 
and scattering transmittances.

\subsection{Aerosol Extinction}
\label{sec:aerosol_ext}

The ratio of stellar flux measured on the ground to that above the atmosphere 
is obtained by integrating over the instrument's spectral response:

\begin{equation}
    \frac{I_{\text{ground}}}{I_{\text{space}}} = \int F(\lambda)\, R(\lambda)\, \boldsymbol{T}_{\text{mol}}(\lambda)\, \exp(-\tau_{\text{aer}}(\lambda)\, A_A)\, \mathrm{d}\lambda
    \label{eq:integral_aerosol}
\end{equation}

where $F(\lambda)$ is the normalised stellar spectral energy distribution, 
averaged over a representative sample of stellar spectra from \citet{Coelho2014} 
spanning the range of stellar types present in the photometric catalogue, 
$R(\lambda)$ is the instrument spectral response, and $\boldsymbol{T}_{\text{mol}}(\lambda)$ 
is the molecular atmospheric transmittance from MODTRAN. Including the molecular 
transmittance in the integration modifies the effective spectral distribution 
of starlight before it interacts with the aerosol layer, making the resulting 
coefficients site-specific. The aerosol extinction coefficients are computed 
by evaluating this integral for three representative values of the \r{A}ngstr\"{o}m 
exponent, $\alpha \in \{0, 1, 2\}$, spanning the physically realistic range for 
natural atmospheric aerosols. In reality, starlight is partially absorbed 
and scattered by molecules before reaching the aerosol layer. This preferentially 
removes shorter wavelengths, modifying the effective spectral distribution of 
starlight that interacts with the aerosols. Omitting this effect (i.e.\ setting 
$\boldsymbol{T}_{\text{mol}}(\lambda) = 1$ in Equation~\ref{eq:integral_aerosol}) 
yields aerosol extinction coefficients that are independent of site altitude:

\begin{equation}
    g_A(A_A, B-V, k_A) = (1 + 0.011\alpha)k_A A_A - 0.023\,\alpha\, k_A A_A(B-V) - 0.0022\,\alpha^2 k_A^2 A_A^2
    \label{eq:aerosol_original}
\end{equation}

When the molecular atmosphere is included, we obtain for CTAO-S0 
(Paranal, 2380~m):

\noindent\textbf{B filter ($\lambda_0 = 440$~nm):}
\begin{equation}
    g_A = (1 + 0.011\alpha)k_A A_A - 0.025\,\alpha\, k_A A_A(B-V) - 0.002366\,\alpha^2 k_A^2 A_A^2
    \label{eq:aerosol_ctas0_B}
\end{equation}

\noindent\textbf{V filter ($\lambda_0 = 540$~nm):}
\begin{equation}
    g_A = (1 + 0.011\alpha)k_A A_A - 0.03365\,\alpha\, k_A A_A(B-V) - 0.003691\,\alpha^2 k_A^2 A_A^2
    \label{eq:aerosol_ctas0_V}
\end{equation}

\noindent\textbf{R filter ($\lambda_0 = 620$~nm):}
\begin{equation}
    g_A = (1 + 0.011\alpha)k_A A_A - 0.014407\,\alpha\, k_A A_A(B-V) - 0.002118\,\alpha^2 k_A^2 A_A^2
    \label{eq:aerosol_ctas0_R}
\end{equation}

The coefficients differ between filters due to the different effective 
wavelengths, but are the same across all three CTAO sites for a given filter. 
Assuming $\alpha = 1$ as the default value introduces a relative uncertainty of 
approximately 1\% in $k_A$ across the range $\alpha \in [0, 2]$, which is 
negligible compared to other systematic uncertainties.

The aerosol airmass $A_A$ in the above equations follows the tropospheric formula, placing all aerosols in the lower atmosphere. Since FRAM measures only the integrated VAOD with no information about the vertical distribution of aerosols, this choice introduces a systematic uncertainty, which is quantified in Section~\ref{sec:configurations} by testing the stratospheric airmass $A_O$ as an alternative.

\section{\r{A}ngstr\"{o}m Exponent}
\label{sec:angstrom}

The \r{A}ngstr\"{o}m exponent describes the wavelength dependence of the 
vertical aerosol optical depth (VAOD). It is defined by the relationship:
\begin{equation}
\label{eq:angstrom_def}
    \alpha = -\frac{\ln(\tau_1/\tau_2)}{\ln(\lambda_1/\lambda_2)}
\end{equation}
where $\tau_1$ and $\tau_2$ are the aerosol optical depths at wavelengths 
$\lambda_1$ and $\lambda_2$, respectively. If the VAOD $\tau_0$ is known at 
a reference wavelength $\lambda_0$, the VAOD at any other wavelength $\lambda$ 
can be obtained from the power-law relation:
\begin{equation}
\label{eq:vaod_to_vaod}
    \tau(\lambda) = \tau_0 \left(\frac{\lambda}{\lambda_0}\right)^{-\alpha}
\end{equation}

The power-law relation in Equation~\ref{eq:angstrom_def} is an 
approximation --- real aerosol optical depth spectra do not follow 
a perfect power law across all wavelengths, particularly for mixed 
aerosol populations or across wide wavelength ranges 
\citep{article, MISHRA2015e00036}.

\subsection{Analysis}
\label{sec:analysis}

The data presented in this work were recorded between 2021 January 1 
and 2022 December 31 at the three CTAO FRAM sites. The pipeline 
configuration used for this analysis is the final configuration 
summarised in Table~\ref{tab:final_config}.

\begin{figure*}
    \centering
    \includegraphics[width=\textwidth]{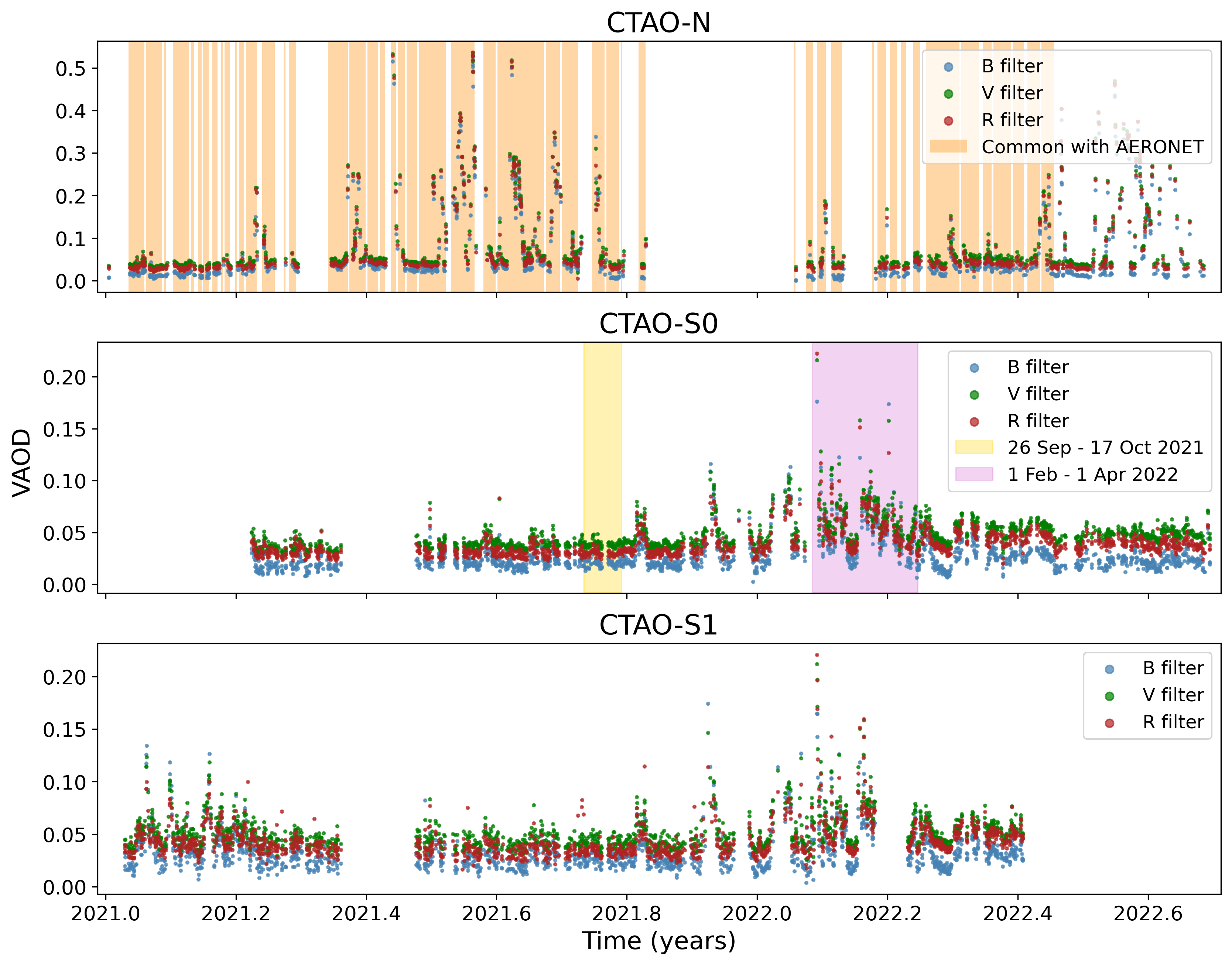}
    \caption{Vertical aerosol optical depth (VAOD) measured in the $B$ 
    (440\,nm), $V$ (540\,nm), and $R$ (620\,nm) filters for the three CTAO 
        sites: CTAO-N (top), CTAO-S0 (middle), and CTAO-S1 (bottom), over the 
    period 2021 January 1 to 2022 December 31. The yellow highlighted region 
    (2021 September 26 to October 17) corresponds to the data period used for 
    the aperture dependence analysis (Section~\ref{sec:aperture_dependence}). 
    The purple highlighted region (2022 February 1 to April 1) corresponds to 
    the data used for the optimum aperture selection 
    (Section~\ref{sec:aperture}).}
    \label{fig:vaod}
\end{figure*}

Figure~\ref{fig:vaod} shows the VAOD time series (preliminary results were 
presented in \citealt{Negi_2025}) measured simultaneously 
in the $B$, $V$, and $R$ filters at all three CTAO sites. The southern 
sites CTAO-S0 and CTAO-S1 show generally stable and low aerosol optical 
depth, with typical VAOD values in the $B$ filter around 0.02--0.05. Both 
southern sites show elevated VAOD towards the end of 2021 and 
into early 2022, which coincides with the volcanic eruption at Hunga 
Tonga on 15 January 2022 \citep{HungaTonga}. 
CTAO-N shows considerably higher and more variable aerosol optical depth, 
with two extended periods of elevated VAOD reaching up to $\sim$0.5 in 
the $B$ filter during mid-2021, consistent with Calima events --- episodic 
intrusions of Saharan dust that are characteristic of the La Palma site.

\begin{figure*}
    \centering
    \includegraphics[width=\textwidth]{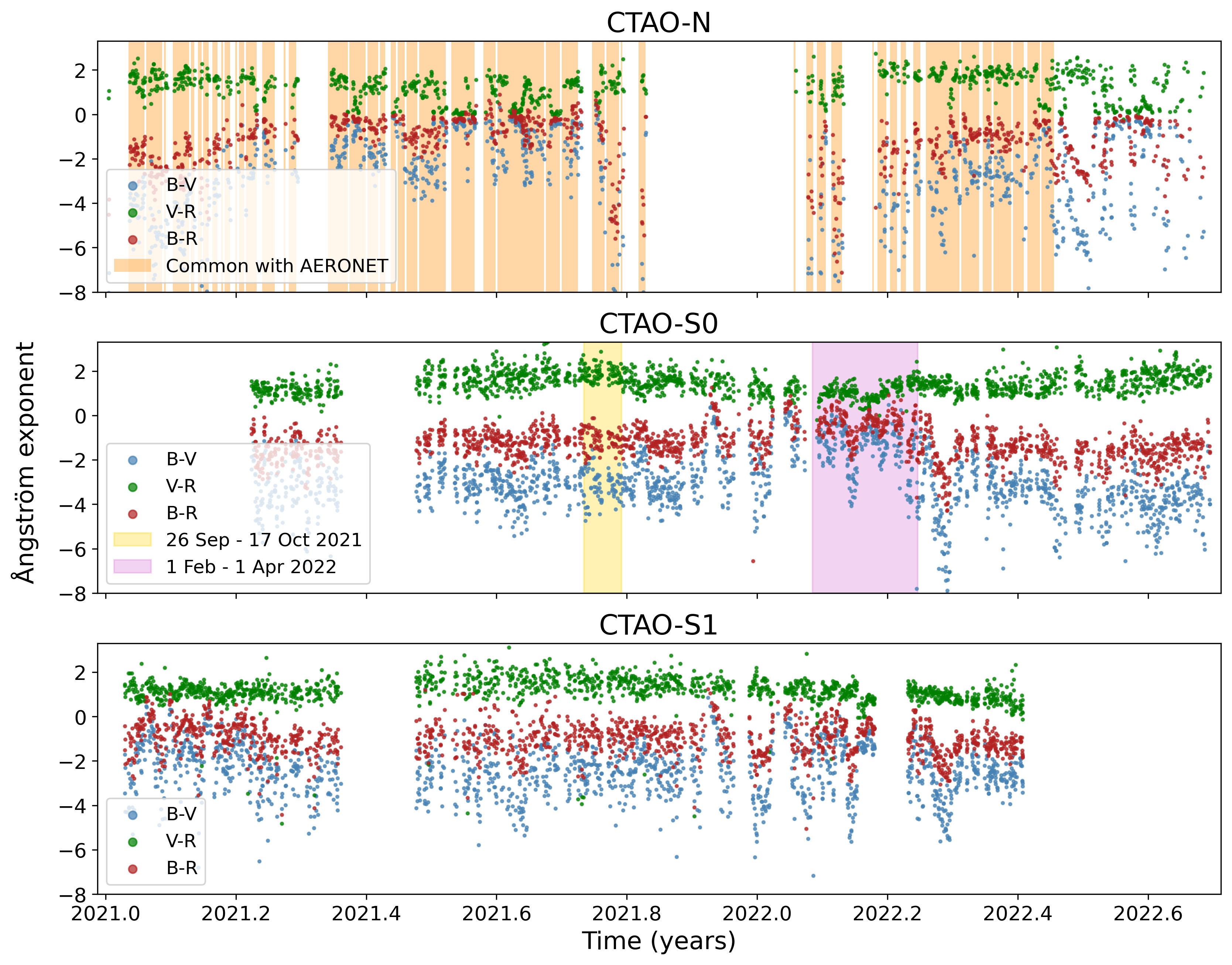}
    \caption{\r{A}ngstr\"{o}m exponent computed from three filter pair 
    combinations: B--V (440--540\,nm), V--R (540--620\,nm), and B--R 
    (440--620\,nm) for CTAO-N (top), CTAO-S0 (middle), and CTAO-S1 (bottom), 
    over the period 2021 January 1 to 2022 December 31. The yellow highlighted 
    region (2021 September 26 to October 17) corresponds to the data period 
    used for the aperture dependence analysis 
    (Section~\ref{sec:aperture_dependence}). The purple highlighted region 
    (2022 February 1 to April 1) corresponds to the data used for the optimum 
    aperture selection (Section~\ref{sec:aperture}).}
    \label{fig:angstrom}
\end{figure*}

Figure~\ref{fig:angstrom} shows the \r{A}ngstr\"{o}m exponent computed 
from the three filter pair combinations. The aerosol optical depth is 
expected to decrease with increasing wavelength, giving the ordering 
$\mathrm{VAOD}_B > \mathrm{VAOD}_V > \mathrm{VAOD}_R$, which would 
produce positive \r{A}ngstr\"{o}m exponent values for all three filter 
pairs. Instead, we observe $\mathrm{VAOD}_V > \mathrm{VAOD}_R > 
\mathrm{VAOD}_B$ across all three sites and the full observing period, 
which causes the B--V and B--R combinations to give negative 
\r{A}ngstr\"{o}m values since VAOD$_B$ is smaller than both VAOD$_V$ 
and VAOD$_R$. The V--R combination gives positive values since 
$\mathrm{VAOD}_V > \mathrm{VAOD}_R$ is the correct relative ordering, however the values 
are small due to the narrow wavelength baseline of this filter pair 
(540--620~nm).

\subsection{Processing Configurations}
\label{sec:configurations}

The $B$-filter VAOD agrees with the previously validated Tycho-2 result 
\citep{Ebr2021}; however, this agreement alone does not establish which 
filter yields the correct VAOD. The inconsistency could arise from the 
$B$ filter, from the $V$ and $R$ filters, or from all three 
simultaneously --- at present this cannot be determined. To investigate 
its origin, we analysed the same dataset under four different 
instrumental configurations. Each configuration modifies one aspect of 
the pipeline configuration relative to the baseline defined in 
Section~\ref{sec:final_config}, allowing us to assess whether the 
observed filter ordering can be attributed to a specific configuration 
choice. The baseline configuration serves as the reference and all 
other configurations are compared against it.

To quantify the impact of each configuration on the derived VAOD, we 
compute for every matched observation the difference 
$\Delta\mathrm{VAOD} = \mathrm{VAOD}_{\mathrm{opt}} - 
\mathrm{VAOD}_{\mathrm{config}}$ in each filter separately. The mean 
and root-mean-square (RMS) of each distribution are evaluated after 
iterative $3\sigma$ clipping.

\subsubsection{Aperture = 5}
\label{sec:config_aper5}

\begin{figure*}
\centering
\includegraphics[width=\textwidth]{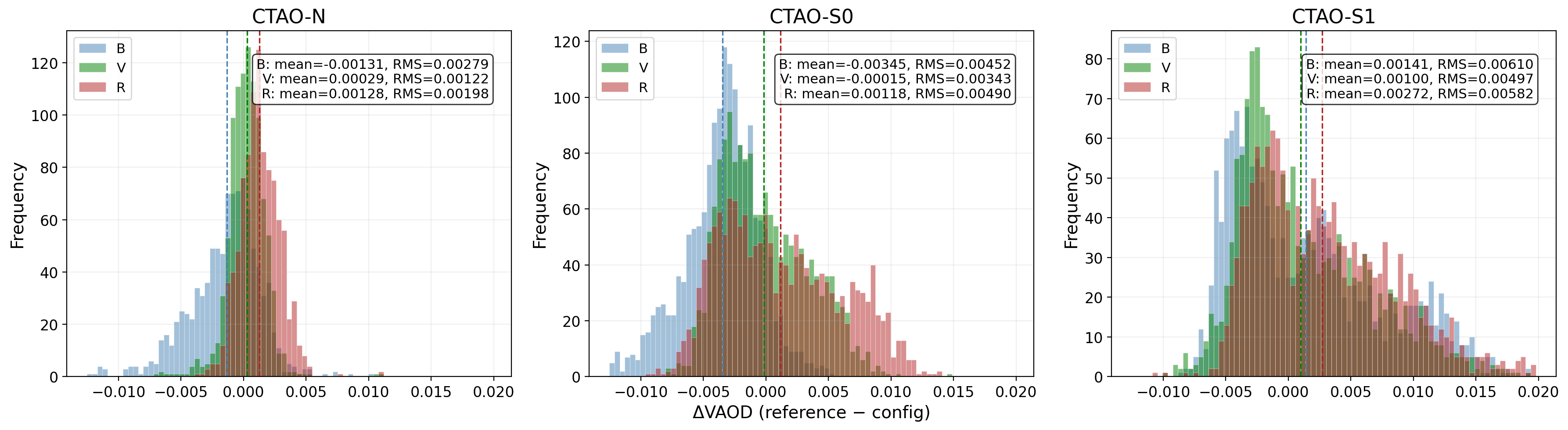}
\caption{Distributions of $\Delta\mathrm{VAOD} = 
\mathrm{VAOD}_\mathrm{opt} - \mathrm{VAOD}_\mathrm{aper5}$ for each 
site (panels) with all three filters ($B$, $V$, $R$) overlaid. The 
dashed lines mark the $3\sigma$-clipped mean for each filter.}
\label{fig:vaod_diff_aperture5}
\end{figure*}

For CTAO-N, the 
aperture selection analysis in Section~\ref{sec:aperture} shows 
a broader and flatter minimum around 4--5~pixels for the $B$ 
filter, making a larger fixed aperture a plausible alternative 
to the optimum values. This configuration fixes the aperture 
radius at 5~pixels uniformly across all filters and all three 
sites, with all other pipeline configuration parameters remaining 
identical to the baseline. The resulting $\Delta\mathrm{VAOD}$ 
distributions are shown in Figure~\ref{fig:vaod_diff_aperture5}.

\subsubsection{Stratospheric Aerosol Airmass}
\label{sec:config_ozone}

\begin{figure*}
\centering
\includegraphics[width=\textwidth]{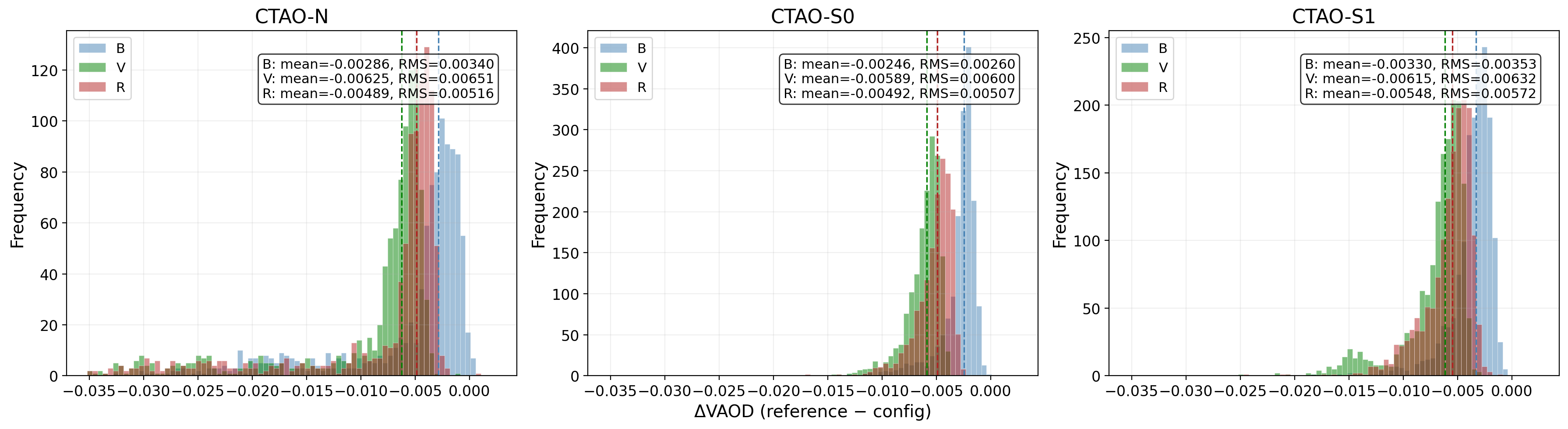}
\caption{Same as Figure~\ref{fig:vaod_diff_aperture5} but for the 
stratospheric airmass configuration. The systematic negative shift 
indicates that the stratospheric airmass configuration yields higher 
VAOD than the baseline, with the effect strongest in the $V$ and 
$R$ filters.}
\label{fig:vaod_diff_ozone}
\end{figure*}

The baseline adopts the tropospheric airmass $A_A$ (Section~\ref{sec:aerosol_ext}). 
This configuration replaces it with the stratospheric airmass 
$A_O$ in the magnitude equation (Equation~\ref{eq:mag}), 
representing the opposite extreme, while leaving the aerosol 
extinction coefficients unchanged. The systematic negative shift 
visible in Figure~\ref{fig:vaod_diff_ozone} across all sites and 
filters is physically expected --- assuming aerosols are at higher 
altitude increases the path length through the aerosol layer for 
a star observed near the horizon, resulting in higher derived 
VAOD. Since the tropospheric assumption gives results more 
consistent with the $B$-filter Tycho-2 validation 
\citep{Ebr2021}, it is retained as the default.

\subsubsection{No Non-Linearity Correction}
\label{sec:config_nonlc}

\begin{figure*}[htbp]
\centering
\includegraphics[width=\textwidth]{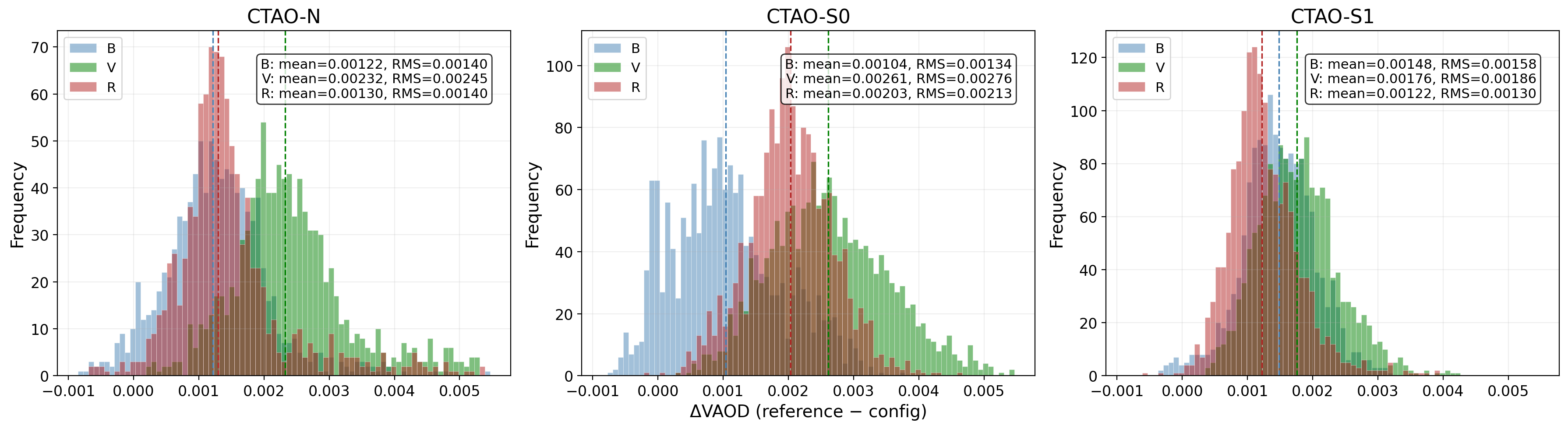}
\caption{Same as Figure~\ref{fig:vaod_diff_aperture5} but with the 
non-linearity correction factor $M$ removed from the magnitude 
equation. The small positive mean offsets indicate that $M$ slightly 
increases the derived VAOD, with the effect most pronounced in the 
$V$ filter.}
\label{fig:vaod_diff_nonlc}
\end{figure*}

The magnitude scaling factor $M$ accounts for any deviation of the 
instrumental magnitude scale from the catalogue magnitude scale, 
including residual effects from CCD non-linearity, flat-fielding, 
and mismatches between the catalogue and instrumental photometric 
systems. Removing $M$ from the magnitude equation 
(Equation~\ref{eq:mag}) therefore tests whether the photometric 
processing is self-consistent --- a small difference between the 
corrected and uncorrected results is a positive indicator of 
well-executed photometry. Of all four configurations, this one 
produces the smallest differences from the baseline, with mean 
offsets between $+0.001$ and $+0.003$ and RMS values below 
$0.003$ across all sites and filters. The resulting 
$\Delta\mathrm{VAOD}$ distributions are shown in 
Figure~\ref{fig:vaod_diff_nonlc}.

\subsubsection{PSF Aperture Correction}
\label{sec:config_psf}

\begin{figure*}[htbp]
\centering
\includegraphics[width=\textwidth]{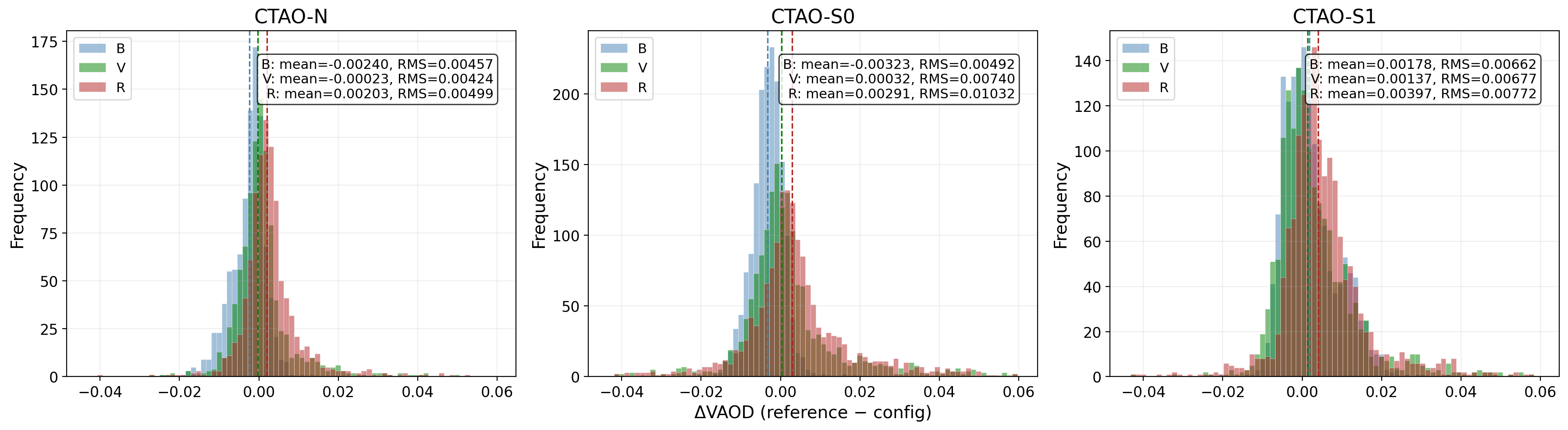}
\caption{Same as Figure~\ref{fig:vaod_diff_aperture5} but with a 
per-star PSF aperture correction applied \citep{Negi2022}. Compared 
to the other configurations, the PSF correction produces the largest 
scatter and most asymmetric distributions.}
\label{fig:vaod_diff_psf}
\end{figure*}

In wide-field aperture photometry, the PSF deviates from a Gaussian 
due to coma aberration as the distance from the optical axis 
increases. At the image periphery the PSF loses rotational symmetry 
and develops asymmetric wings, causing a larger fraction of stellar 
flux to fall outside the fixed aperture compared to stars near the 
centre. The aperture correction accounts for this position-dependent 
flux loss by estimating, for each star individually, the ratio of 
flux inside the aperture to the total flux using a PSF model 
extracted with PSFEx, as described in detail in \citet{Negi2022}. 
Of the four configurations, the PSF correction produces the largest 
scatter and is the least stable, with RMS values reaching up to 
$0.010$ in the $R$ filter at CTAO-S0, reflecting the sensitivity 
of the correction to the PSF model quality across the wide field 
of view. The resulting $\Delta\mathrm{VAOD}$ distributions are 
shown in Figure~\ref{fig:vaod_diff_psf}.

None of the four configurations resolves the systematic filter 
ordering $\mathrm{VAOD}_V > \mathrm{VAOD}_R > \mathrm{VAOD}_B$, 
confirming that it cannot be attributed to any single pipeline 
configuration choice tested here. Among the four, the no non-linearity correction 
configuration produces the smallest systematic differences from the 
baseline, while the PSF correction introduces the largest scatter 
and is the least suitable for operational use. This motivates the 
intercept correction method described in Section~\ref{sec:intercept}.

\section{Intercept Method}
\label{sec:intercept}

When the VAOD measured in one filter is plotted against the VAOD measured in
another, the relationship is expected to pass through the origin. However, in
practice a linear fit to paired VAOD measurements,
\begin{equation}
\label{eq:vaod_linear}
    \tau_y = A \cdot \tau_x + B,
\end{equation}
yields a non-zero intercept $B$. This intercept, if left uncorrected,
biases the derived \r{A}ngstr\"{o}m exponent. The intercept correction
accounts for this by subtracting $B$ from $\tau_y$ --- the VAOD of the
filter on the y-axis of the linear fit --- before computing $\alpha$,
as shown in Equation~\ref{eq:angstrom_corrected}. The corrected
\r{A}ngstr\"{o}m exponent is then:
\begin{equation}
\label{eq:angstrom_corrected}
    \alpha = -\frac{\ln\!\left(\tau_x \,/\, (\tau_y - B)\right)}{\ln(\lambda_x / \lambda_y)}.
\end{equation}

The intercept correction relies on a linear relationship between the VAODs of two filters, which holds only when the \r{A}ngstr\"{o}m exponent is approximately constant across the fitted data. Since both
VAOD measurements carry uncertainties, we use Orthogonal Distance
Regression (ODR) \citep{Boggs1990} to fit the linear relation,
accounting for measurement uncertainties on both axes simultaneously.

\begin{figure*}[htbp]
    \centering
    \includegraphics[width=\textwidth]{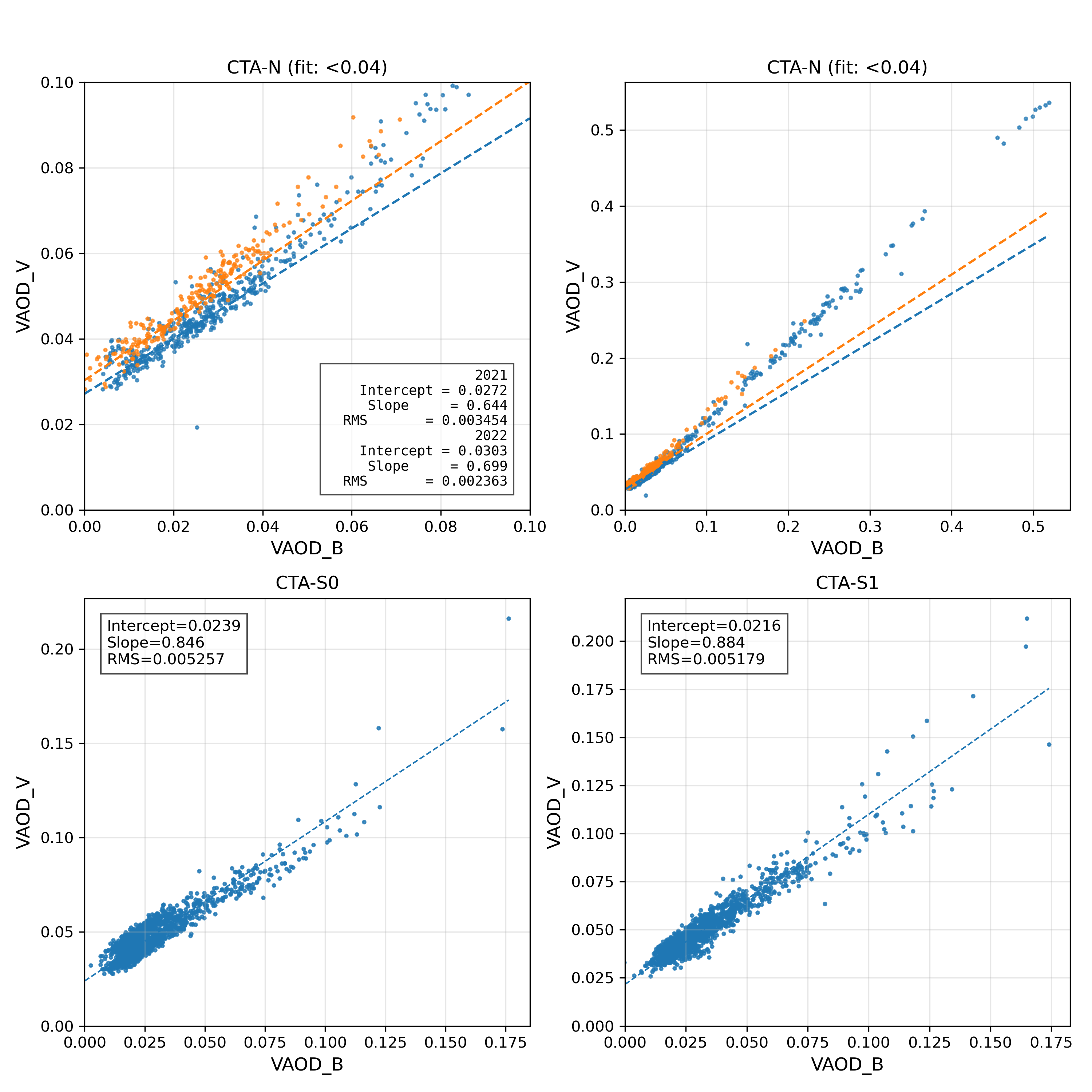}
    \caption{Scatter plots of VAOD$_B$ vs VAOD$_V$ for CTAO-N (top row) and
    CTAO-S0/S1 (bottom row), showing the B--V filter pair as a representative
    example. Linear fits are performed using ODR \citep{Boggs1990} to account
    for measurement uncertainties in both variables. Top-left: CTAO-N with
    VAOD~$<~0.1$ displayed. Top-right: CTAO-N full dataset. In both CTAO-N
    panels the data are restricted to dates common with AERONET and coloured
    by year, and separate fits are shown for 2021 (blue) and 2022 (orange); each fit uses
    scans with VAOD$_B < 0.04$. Bottom-left: CTAO-S0 full dataset.
    Bottom-right: CTAO-S1 full dataset. The non-zero intercepts of the fitted
    lines represent the filter-to-filter systematic offset corrected by the
    intercept correction.}
    \label{fig:vaod_scatter}
\end{figure*}

Figure~\ref{fig:vaod_scatter} shows the scatter of VAOD$_V$ against
VAOD$_B$ for the B--V filter pair as a representative example, with
the fit parameters and the RMS of the scatter around the fitted line
summarised in Table~\ref{tab:intercept_rms} for all sites and filter
pairs. At CTAO-S0 and CTAO-S1, the aerosol type is relatively stable
and the data follow an almost single linear relationship across the
full range of observed VAOD values. At CTAO-N, the aerosol population
is not stable but exhibits a pronounced bimodal structure --- two
distinct linear relationships are naturally visible in the VAOD
scatter plot. If the aerosol type varied continuously and randomly,
no stable linear relationship between filter pairs would exist and the
intercept correction would not be applicable.

For CTAO-N the intercept is derived from the low aerosol loading
subset, where the data are most densely clustered and the linear
relationship is most reliably determined, and is then applied to the
full dataset. The selection is made at the level of individual scans:
a scan is retained for the fit if its VAOD$_B$ lies below 0.04, and
the corresponding $V$ and $R$ measurements are included regardless of
their own values, since all three filters observe the same field
simultaneously. The CTAO-N analysis is further restricted to nights
for which contemporaneous AERONET measurements exist, and the fit is
performed separately for 2021 and 2022; the motivation for this
year-wise treatment is discussed in Section~\ref{sec:results_ctan}.
For CTAO-S0 and CTAO-S1 the intercept is derived from all available
data.

\begin{table}[htbp]
\centering
\caption{ODR fit parameters $\tau_y = A \cdot \tau_x + B$ for each site
and filter pair. The RMS is the root-mean-square scatter of the data
around the fitted line. For CTAO-N the fit uses scans with
VAOD$_B < 0.04$, restricted to dates common with AERONET and computed
separately for each year; for CTAO-S0 and CTAO-S1 all available data
are used. Only three of the six filter pairs are listed, as the reverse 
pair follows exactly from $A \to 1/A$ and $B \to -B/A$.}
\setlength{\tabcolsep}{4pt}
\begin{tabular}{lcccccc}
\hline
 & \multicolumn{3}{c}{CTAO-N 2021} & \multicolumn{3}{c}{CTAO-N 2022} \\
\cline{2-4} \cline{5-7}
Pair & $B$ & $A$ & RMS & $B$ & $A$ & RMS \\
\hline
B--V & $0.026$    & 0.753 & 0.0034 & $0.029$    & 0.830 & 0.0025 \\
V--R & $-0.006$   & 0.954 & 0.0018 & $-0.002$   & 0.829 & 0.0019 \\
R--B & $-0.026$   & 1.393 & 0.0056 & $-0.032$   & 1.453 & 0.0051 \\
\hline
 & \multicolumn{3}{c}{CTAO-S0} & \multicolumn{3}{c}{CTAO-S1} \\
\cline{2-4} \cline{5-7}
Pair & $B$ & $A$ & RMS & $B$ & $A$ & RMS \\
\hline
B--V & $0.024$    & 0.846 & 0.005 & $0.022$    & 0.884 & 0.005 \\
V--R & $-0.006$   & 0.951 & 0.003 & $-0.006$   & 0.983 & 0.005 \\
R--B & $-0.021$   & 1.244 & 0.007 & $-0.017$   & 1.151 & 0.009 \\
\hline
\end{tabular}
\label{tab:intercept_rms}
\end{table}

The data follow the fitted line tightly (Figure~\ref{fig:vaod_scatter}
shows the B--V pair as an example), particularly for the CTAO-N low
VAOD population. For this population the RMS values in
Table~\ref{tab:intercept_rms} range from 0.0018 to 0.0056 across the
three filter pairs and both years, all smaller than the quoted FRAM
measurement precision of 0.008 optical depths \citep{Ebr2021}. The RMS
values at CTAO-S0 and CTAO-S1 are also small. This unexpectedly small
scatter implies that the \r{A}ngstr\"{o}m exponent must be nearly
constant within each aerosol population.

The intercept values show a degree of consistency across sites. 
For CTAO-S0 and CTAO-S1 the intercept is derived from all 
available data combined, while for CTAO-N separate fits are 
performed for 2021 and 2022. For the B--V pair the intercept 
lies between 0.022 and 0.029 across all three sites and both 
CTAO-N years. The V--R intercept is $-0.006$ at CTAO-S0, 
CTAO-S1 and CTAO-N in 2021, with a small deviation at CTAO-N 
in 2022 ($-0.002$). The R--B intercept is consistent in sign 
throughout ($-0.017$ to $-0.032$). This cross-site consistency 
suggests that the intercept originates from a systematic effect 
common to all three FRAM telescopes and their processing 
pipelines, rather than from site-specific atmospheric conditions. 
The nature of this systematic effect remains an open question.

The \r{A}ngstr\"{o}m exponent can also be derived directly from 
the slope of the VAOD scatter plot. Substituting 
$\tau_y - B = A\,\tau_x$ into Equation~\ref{eq:angstrom_corrected} 
gives
\begin{equation}
\label{eq:alpha_slope}
    \alpha = \frac{\ln A}{\ln(\lambda_x / \lambda_y)},
\end{equation}
which is manifestly invariant under exchange of the two filters. For
the CTAO-N low aerosol loading population this yields
$\alpha = 1.39$ (B--V), $0.34$ (V--R) and $0.97$ (R--B) in 2021, and
$0.91$, $1.35$ and $1.09$ respectively in 2022. For all six filter pairs and both years, the slope-derived values agree with the corresponding point-by-point means in the fine regime to within 0.15 (Table~\ref{tab:yearwise}), providing an independent check on the correction. The V--R pair is the least stable between the two years: its slope-derived exponent changes from 0.34 to 1.35, against 1.39 to 0.91 for B--V and 0.97 to 1.09 for R--B. Its two effective wavelengths are separated by only 80~nm, so $\ln(\lambda_x/\lambda_y)$ is small and any change in the fitted slope is amplified in $\alpha$ by a factor of 7.2, compared with 4.9 for B--V and 2.9 for B--R. The same amplification applies to the intercept
subtraction, since $B$ is subtracted from a $\tau_R$ of comparable
magnitude to $B$ itself. Results derived from the V--R pair should
therefore be treated with corresponding caution.

\subsection{CTAO-N, La Palma and AERONET Comparison}
\label{sec:results_ctan}

The comparison presented here was motivated by \citet{Schmuckermaier2023},
who required the \r{A}ngstr\"{o}m exponent to extrapolate aerosol
extinction from the LIDAR wavelength of 532~nm to the wavelengths
relevant for Cherenkov light detection at the same site, and derived
it from AERONET sun-photometer data at the Roque de los Muchachos
Observatory. They found a bimodal \r{A}ngstr\"{o}m exponent
distribution corresponding to Calima dust intrusions
($\alpha \approx 0.32 \pm 0.21$) and clean background conditions
($\alpha \approx 1.45 \pm 0.46$). Since the FRAM telescope at CTAO-N
is located at the same site, the same two populations are expected
in the FRAM data.

To assess whether the observed \r{A}ngstr\"{o}m exponent distributions
are physically meaningful, we used AERONET Level~2.0 sun-photometer
measurements from the Roque de los Muchachos Observatory, located
approximately 400~m from the FRAM telescope and 30~m lower in
altitude, covering the same 2021--2022 period. For the
\r{A}ngstr\"{o}m exponent calculation, we selected AERONET wavelength
pairs closest to the FRAM effective wavelengths of 440, 540, and
620~nm: 440--500~nm as the B--V equivalent, 500--675~nm as the V--R
equivalent, and 440--675~nm as the B--R equivalent.

\subsubsection{Temporal matching of the two datasets}

The two instruments never measure simultaneously, as AERONET 
operates during the day and FRAM at night, with neither operating 
continuously. Comparing distributions accumulated over the full 
two-year period would therefore average each instrument over a 
different set of atmospheric conditions, and any difference between 
them could reflect the sampling rather than the atmosphere. All 
comparisons presented in this section are consequently restricted 
to the 300 calendar dates on which both instruments obtained usable 
measurements; these dates are indicated in Figure~\ref{fig:vaod}. 
This yields 15{,}759 AERONET direct-sun measurements and 811 FRAM 
scans at CTAO-N.

Restricting to common dates does not make the two years equivalent. The 
common dates in 2021 span January to October, whereas those in 
2022 end in June, so the late-summer months --- precisely when 
Calima intrusions are most frequent at this site --- are represented 
in 2021 and absent in 2022. Combining the two years into a single 
distribution would therefore weight the aerosol populations unevenly 
and obscure this imbalance. We accordingly perform the entire 
analysis separately for each year: the intercept is fitted and 
$\alpha$ computed independently within 2021 and 2022, and the 
resulting distributions are combined only at the final stage where 
a two-year value is quoted.

\subsubsection{Aerosol regimes}

\begin{figure*}[htbp]
    \centering
    \includegraphics[width=\textwidth]{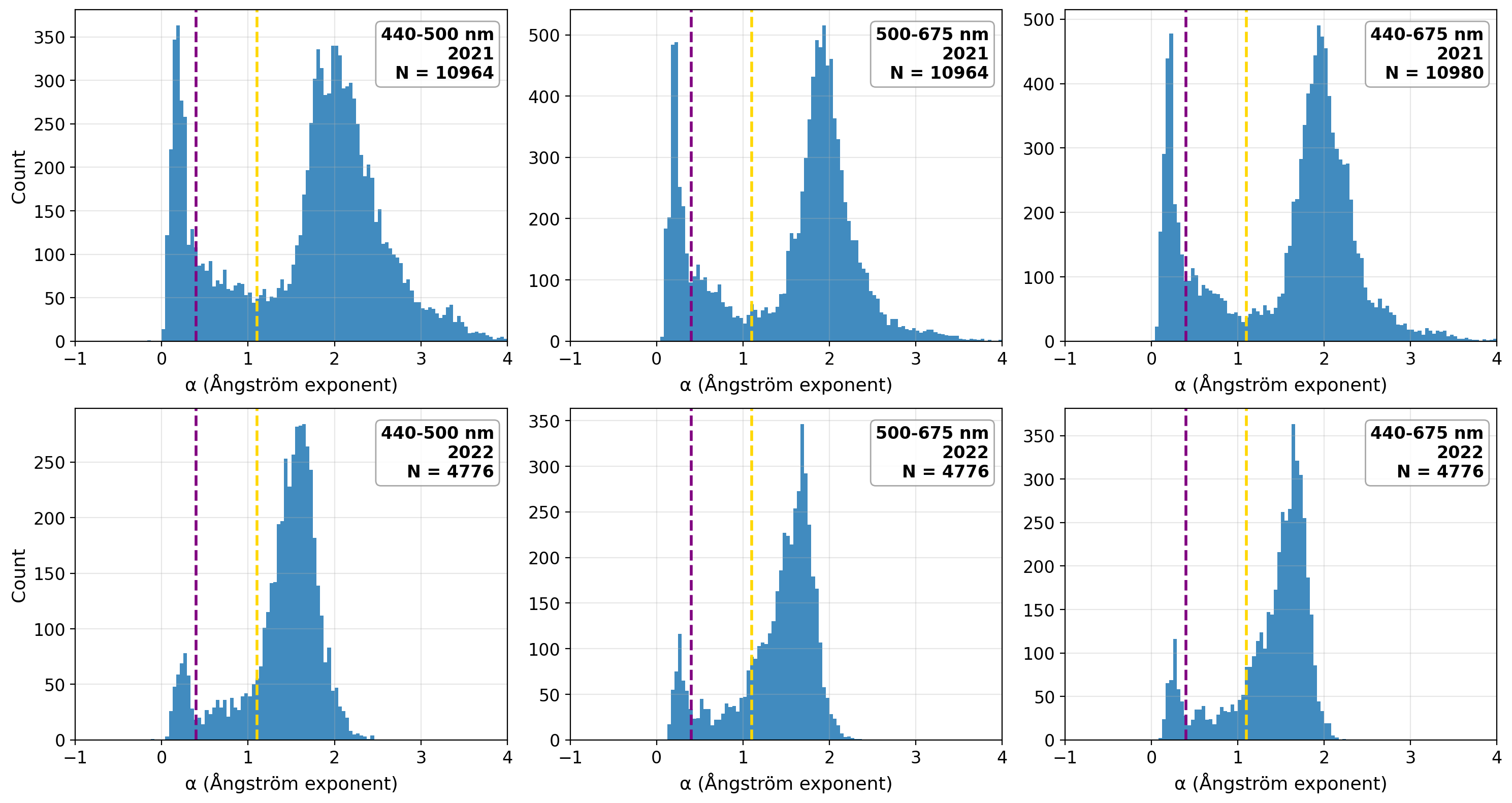}
    \caption{\r{A}ngstr\"{o}m exponent distributions from AERONET
    measurements at Roque de los Muchachos for three wavelength pairs
    selected to match FRAM filter combinations: 440--500~nm (B--V
    equivalent), 500--675~nm (V--R equivalent), and 440--675~nm (B--R
    equivalent), shown separately for 2021 (top row) and 2022 (bottom
    row). Only dates common with FRAM are included. Purple and gold
    dashed lines at $\alpha = 0.4$ and $\alpha = 1.1$ define Region~1
    (coarse aerosols), Region~2 (transition) and Region~3 (fine
    aerosols). Each panel is scaled independently.}
    \label{fig:aeronet_fram_pairs}
\end{figure*}

The AERONET \r{A}ngstr\"{o}m exponent distributions for the three 
wavelength pairs are shown in Figure~\ref{fig:aeronet_fram_pairs}. 
For simplicity, the 2021 and 2022 data are combined when defining 
the region boundaries, so that a single set of regions applies 
throughout; the mean \r{A}ngstr\"{o}m exponent and population 
fractions within each region are computed separately for each year. 
The distributions separate naturally into three populations based 
on aerosol particle size: coarse-dominated aerosols 
($\alpha < 0.4$, Region~1), a transition regime 
($0.4 \leq \alpha < 1.1$, Region~2), and fine-dominated aerosols 
($\alpha \geq 1.1$, Region~3). Over the combined two-year period, 
Region~1 accounts for approximately 15\% of observations and 
corresponds to episodic Saharan dust intrusions (Calima events), 
Region~2 for $\sim$11\%, and Region~3 for $\sim$75\%, representing 
clean background conditions dominated by fine-mode aerosols. The 
year-wise panels of Figure~\ref{fig:aeronet_fram_pairs} show the 
effect of the seasonal coverage difference directly: the coarse 
population is substantially better represented in 2021 than in 2022.

The mean AOD values in each region are summarised in
Table~\ref{tab:aod_means_by_region}. Region~1 shows high mean AOD
($\tau_{440} = 0.250$) with a weak wavelength dependence, consistent
with large coarse-mode particles. Region~3 shows low mean AOD
($\tau_{440} = 0.039$) with a stronger wavelength dependence
($\tau_{675} = 0.018$), consistent with fine-mode aerosols.  The AERONET AOD
distribution at 440~nm --- the wavelength coinciding with the FRAM $B$
filter effective wavelength --- reveals natural boundaries at 0.04 and
0.1. These were adopted as thresholds to divide the FRAM dataset by
scan into three subsets based on VAOD$_B$, since each scan covers all
three filters simultaneously: fine (VAOD$_B < 0.04$), moderate
($0.04 \leq \mathrm{VAOD}_B < 0.1$), and coarse
(VAOD$_B \geq 0.1$). A lower bound of VAOD$_B \geq 0.02$ is additionally
imposed when computing $\alpha$, below which the intercept-corrected
optical depth becomes comparable to the correction itself.

\begin{table*}[htbp]
\centering
\caption{Mean AOD values by \r{A}ngstr\"{o}m exponent region for AERONET
data at Roque de los Muchachos, restricted to dates common with FRAM.}
\begin{tabular}{lcccccc}
\hline
Wavelength & \multicolumn{2}{c}{Region 1 ($\alpha < 0.4$)} &
\multicolumn{2}{c}{Region 2 ($0.4 \leq \alpha < 1.1$)} &
\multicolumn{2}{c}{Region 3 ($\alpha \geq 1.1$)} \\
\cline{2-3} \cline{4-5} \cline{6-7}
Pair (nm) & $\lambda_1$ & $\lambda_2$ & $\lambda_1$ & $\lambda_2$ &
$\lambda_1$ & $\lambda_2$ \\
\hline
440--500 & 0.250 & 0.244 & 0.108 & 0.099 & 0.039 & 0.031 \\
500--675 & 0.244 & 0.229 & 0.099 & 0.083 & 0.031 & 0.018 \\
440--675 & 0.250 & 0.229 & 0.108 & 0.083 & 0.039 & 0.018 \\
\hline
\end{tabular}
\label{tab:aod_means_by_region}
\end{table*}

\subsubsection{Year-wise comparison}

\begin{figure*}[htbp]
    \centering
    \includegraphics[width=\textwidth]{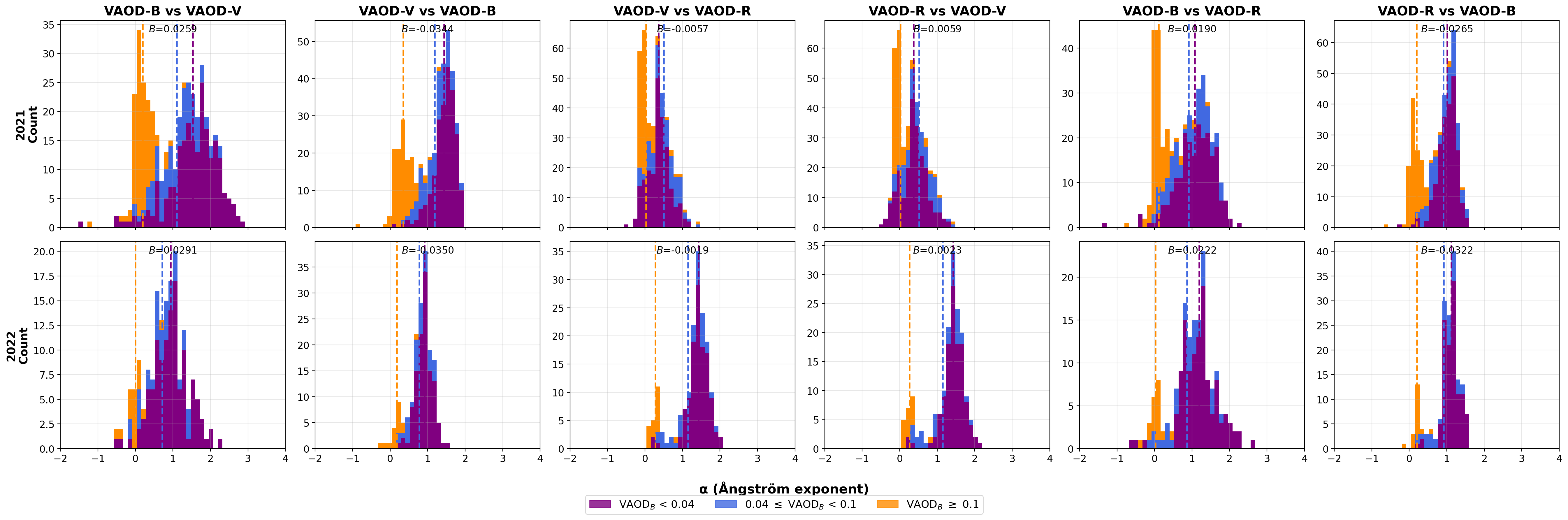}
    \caption{\r{A}ngstr\"{o}m exponent distributions at CTAO-N for the
    optimum aperture configuration, computed separately for 2021 (top
    row) and 2022 (bottom row) across six wavelength pairs. Each
    observation is assigned to one of three aerosol loading regimes
    based on VAOD$_B$: fine ($0.02 \leq \mathrm{VAOD}_B < 0.04$,
    purple), moderate ($0.04 \leq \mathrm{VAOD}_B < 0.1$, blue), and
    coarse (VAOD$_B \geq 0.1$, orange). Dashed vertical lines mark the
    mean $\alpha$ of each regime. The intercept $B$ fitted for that year
    and pair is annotated in each panel.}
    \label{fig:cta-n_by_year}
\end{figure*}

Figure~\ref{fig:cta-n_by_year} shows the FRAM \r{A}ngstr\"{o}m exponent
distributions for the optimum aperture configuration, computed
separately for each year with its own intercept. Table~\ref{tab:yearwise}
compares three independent estimates of $\alpha$ for each year: the
AERONET mean, the value derived from the ODR slope
(Equation~\ref{eq:alpha_slope}), and the mean of the point-by-point
intercept-corrected values in the fine regime.

\begin{table*}[htbp]
\centering
\caption{\r{A}ngstr\"{o}m exponent at CTAO-N for the optimum aperture
configuration, compared year by year against AERONET. The AERONET and
slope-derived values (Equation~\ref{eq:alpha_slope}) are identical for
reversed filter pairs and are therefore given once, centred across each
pair. The intercept-corrected value is the mean of the point-by-point
$\alpha$ in the fine regime ($0.02 \leq \mathrm{VAOD}_B < 0.04$). AERONET
values use the corresponding wavelength pairs.}
\begin{tabular}{llcccccc}
\hline
Year & Estimate & B--V & V--B & V--R & R--V & B--R & R--B \\
\hline
2021
 & AERONET             & \multicolumn{2}{c}{1.651} & \multicolumn{2}{c}{1.515} & \multicolumn{2}{c}{1.560} \\
 & ODR slope           & \multicolumn{2}{c}{1.388} & \multicolumn{2}{c}{0.340} & \multicolumn{2}{c}{0.966} \\
 & Intercept-corrected & 1.535 & 1.443 & 0.367 & 0.371 & 1.081 & 1.017 \\
\hline
2022
 & AERONET             & \multicolumn{2}{c}{1.381} & \multicolumn{2}{c}{1.375} & \multicolumn{2}{c}{1.377} \\
 & ODR slope           & \multicolumn{2}{c}{0.909} & \multicolumn{2}{c}{1.354} & \multicolumn{2}{c}{1.089} \\
 & Intercept-corrected & 0.946 & 0.920 & 1.428 & 1.432 & 1.193 & 1.127 \\
\hline
\end{tabular}
\label{tab:yearwise}
\end{table*}

The slope-derived and intercept-corrected values agree to within 0.15 for all six pairs in both years, confirming internal consistency between the two FRAM estimates. The agreement with
AERONET is good for B--V and B--R in 2021 and for B--R in 2022. The
B--V value falls in both datasets between 2021 and 2022, though the
FRAM decrease is roughly twice as large as the AERONET one.

The V--R pair behaves differently, moving from 0.37 in 2021 to 1.43 in
2022 while the corresponding AERONET value remains near 1.4 in both
years. The underlying VAOD$_R$/VAOD$_V$ ratio in the fine regime
changes by only 5\% between the two years, from 0.824 to 0.780, which
is consistent with the change in seasonal coverage; given the
amplification factor discussed in Section~\ref{sec:intercept}, this
small change in the measured ratio is sufficient to produce the large
apparent shift in $\alpha$. The V--R result should therefore not be
read as evidence of a change in aerosol properties between the two
years.

\subsubsection{Aerosol populations across configurations}

\begin{figure*}[htbp]
\centering
\includegraphics[width=\textwidth]{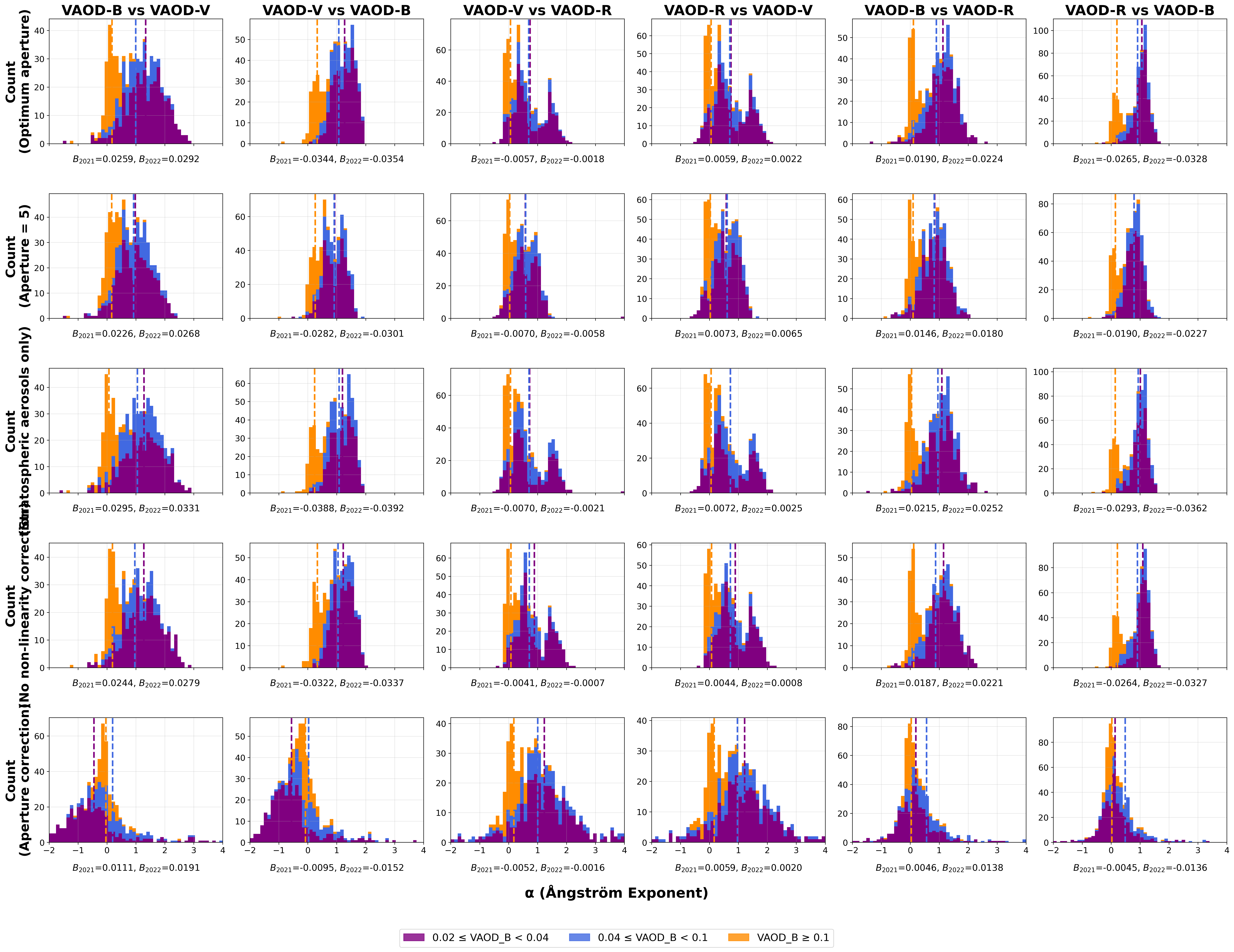}
\caption{\r{A}ngstr\"{o}m exponent ($\alpha$) distributions for CTAO-N
across five instrumental configurations (rows) and six wavelength pairs
(columns), restricted to dates common with AERONET. Each observation is
assigned to one of three aerosol loading regimes based on VAOD$_B$:
fine ($0.02 \leq \mathrm{VAOD}_B < 0.04$, purple), moderate
($0.04 \leq \mathrm{VAOD}_B < 0.1$, blue), and coarse
(VAOD$_B \geq 0.1$, orange). Dashed vertical lines mark the mean
$\alpha$ for each regime. The intercept fitted for each year is annotated beneath each panel.}
\label{fig:cta-n}
\end{figure*}

Figure~\ref{fig:cta-n} shows the \r{A}ngstr\"{o}m exponent
distributions for all five configurations, and
Tables~\ref{tab:stats_ctan_fine} and~\ref{tab:stats_ctan_coarse}
summarise the fine and coarse regimes numerically. 

\begin{table*}[htbp]
\centering
\caption{\r{A}ngstr\"{o}m exponent statistics for CTAO-N in the fine
aerosol regime ($0.02 \leq \mathrm{VAOD}_B < 0.04$), corresponding to
clean maritime background conditions. Each configuration is processed
separately for 2021 and 2022 and the two years combined only at the
final stage. The AERONET row gives the mean over Region~3
($\alpha \geq 1.1$) on the same dates; the ODR slope row follows from
Equation~\ref{eq:alpha_slope} for the optimum aperture configuration,
fitted on scans with $\mathrm{VAOD}_B < 0.04$. Both are identical for
reversed pairs and are therefore given once across each pair.}
\setlength{\tabcolsep}{4pt}
\begin{tabular}{lcccccccccccc}
\hline
 & \multicolumn{2}{c}{B--V} & \multicolumn{2}{c}{V--B} &
\multicolumn{2}{c}{V--R} & \multicolumn{2}{c}{R--V} &
\multicolumn{2}{c}{B--R} & \multicolumn{2}{c}{R--B} \\
\cline{2-3} \cline{4-5} \cline{6-7}
\cline{8-9} \cline{10-11} \cline{12-13}
AERONET (Region 3)        & \multicolumn{4}{c}{1.96} & \multicolumn{4}{c}{1.83} & \multicolumn{4}{c}{1.87} \\
ODR slope (optimum aper.) & \multicolumn{4}{c}{1.15} & \multicolumn{4}{c}{0.85} & \multicolumn{4}{c}{1.03} \\
\hline
 & Mean & RMS & Mean & RMS & Mean & RMS & Mean & RMS &
   Mean & RMS & Mean & RMS \\
\hline
Optimum aperture            & 1.33 & 0.67 & 1.26 & 0.37 & 0.74 & 0.59 & 0.74 & 0.60 & 1.12 & 0.50 & 1.06 & 0.26 \\
Aperture = 5                & 0.96 & 0.61 & 0.92 & 0.39 & 0.58 & 0.44 & 0.58 & 0.44 & 0.83 & 0.47 & 0.79 & 0.28 \\
Stratospheric aerosols only & 1.27 & 0.70 & 1.19 & 0.38 & 0.72 & 0.66 & 0.71 & 0.65 & 1.08 & 0.53 & 1.00 & 0.25 \\
No non-linearity correction & 1.27 & 0.64 & 1.22 & 0.36 & 0.89 & 0.56 & 0.89 & 0.57 & 1.14 & 0.48 & 1.09 & 0.25 \\
Aperture correction (PSF)   & $-0.41$ & 1.03 & $-0.52$ & 0.82 & 1.18 & 0.97 & 1.16 & 0.96 & 0.19 & 0.69 & 0.13 & 0.59 \\
\hline
\end{tabular}
\par\smallskip\noindent{\textit{Note.} Number of FRAM observations: $N = 337$ for the optimum
aperture configuration (219 in 2021, 118 in 2022), varying by a few per
cent between configurations.}
\label{tab:stats_ctan_fine}
\end{table*}

\begin{table*}[htbp]
\centering
\caption{\r{A}ngstr\"{o}m exponent statistics for CTAO-N in the coarse
aerosol regime ($\mathrm{VAOD}_B \geq 0.1$), corresponding to Calima
dust intrusion periods. Columns and processing as in
Table~\ref{tab:stats_ctan_fine}; the AERONET row gives the mean over
Region~1 ($\alpha < 0.4$) and the ODR slope row is fitted on scans with
$\mathrm{VAOD}_B \geq 0.1$. Both are identical for reversed pairs and
are therefore given once across each pair.}
\setlength{\tabcolsep}{4pt}
\begin{tabular}{lcccccccccccc}
\hline
 & \multicolumn{2}{c}{B--V} & \multicolumn{2}{c}{V--B} &
\multicolumn{2}{c}{V--R} & \multicolumn{2}{c}{R--V} &
\multicolumn{2}{c}{B--R} & \multicolumn{2}{c}{R--B} \\
\cline{2-3} \cline{4-5} \cline{6-7}
\cline{8-9} \cline{10-11} \cline{12-13}
AERONET (Region 1)        & \multicolumn{4}{c}{0.21} & \multicolumn{4}{c}{0.23} & \multicolumn{4}{c}{0.23} \\
ODR slope (optimum aper.) & \multicolumn{4}{c}{0.00} & \multicolumn{4}{c}{$-0.10$} & \multicolumn{4}{c}{$-0.04$} \\
\hline
 & Mean & RMS & Mean & RMS & Mean & RMS & Mean & RMS &
   Mean & RMS & Mean & RMS \\
\hline
Optimum aperture            & 0.18 & 0.28 & 0.33 & 0.25 & 0.06 & 0.26 & 0.05 & 0.27 & 0.11 & 0.25 & 0.20 & 0.23 \\
Aperture = 5                & 0.16 & 0.29 & 0.26 & 0.27 & 0.04 & 0.24 & 0.03 & 0.25 & 0.10 & 0.26 & 0.15 & 0.24 \\
Stratospheric aerosols only & 0.07 & 0.27 & 0.24 & 0.25 & 0.06 & 0.26 & 0.05 & 0.27 & 0.04 & 0.24 & 0.14 & 0.22 \\
No non-linearity correction & 0.19 & 0.28 & 0.33 & 0.25 & 0.08 & 0.27 & 0.07 & 0.27 & 0.12 & 0.25 & 0.22 & 0.23 \\
Aperture correction (PSF)   & $-0.02$ & 0.39 & $-0.06$ & 0.36 & 0.17 & 0.36 & 0.15 & 0.37 & 0.04 & 0.32 & 0.03 & 0.31 \\
\hline
\end{tabular}
\par\smallskip\noindent{\textit{Note.} Number of FRAM observations: $N = 128$ for the optimum
aperture configuration, of which only 16 fall in 2022 --- the common
dates in that year end in June and so exclude the late-summer months
when Calima intrusions are most frequent. The 2022 contribution to this
regime should therefore be regarded as poorly sampled.}
\label{tab:stats_ctan_coarse}
\end{table*}

The B--V and B--R filter pairs show a pronounced bimodal structure. The
coarse regime is concentrated at lower $\alpha$ values, corresponding to
Region~1 in the AERONET data and consistent with Calima dust dominated
by large coarse-mode particles. The fine regime is concentrated at
higher $\alpha$ values, corresponding to Region~3 and consistent with
fine-mode maritime aerosols \citep{Schmuckermaier2023}. The moderate
regime bridges the two populations. The V--R filter pair shows a
narrower distribution for all three regimes, since the two effective
wavelengths are close together and the aerosol extinction differs
little between the two filters.

The bimodal structure is consistent across four of the five
instrumental configurations for both the B--V and B--R pairs,
confirming that the result is robust and not an artifact of any
particular pipeline configuration choice. The PSF correction
configuration is an exception --- it produces broader distributions
with larger scatter that obscure the bimodal structure, and drives the
B--V mean to negative values, indicating that the PSF correction as
currently implemented introduces additional noise.

\subsubsection{Statistical and systematic uncertainties}

Two distinct uncertainties affect the mean \r{A}ngstr\"{o}m exponents
reported above. The statistical uncertainty on each mean is the standard
error, $\sigma/\sqrt{N}$, where $\sigma$ is the sample standard
deviation of the individual $\alpha$ values within a regime. The
systematic uncertainty associated with the choice of pipeline
configuration is estimated from half the range of the mean across
configurations. Both quantities are evaluated separately for each year,
since the seasonal coverage of the two years differs and pooling them
would conflate a sampling difference with a measurement uncertainty.
They are listed in Table~\ref{tab:uncertainties}.

\begin{table*}[htbp]
\centering
\caption{Statistical and systematic uncertainties on the mean
\r{A}ngstr\"{o}m exponent at CTAO-N, evaluated separately for each year.
The standard error is quoted for the optimum aperture configuration; the
systematic term is half the range of the mean across configurations,
excluding the PSF correction.}
\begin{tabular}{llllcccccc}
\hline
Regime & Year & Quantity & B--V & V--B & V--R & R--V & B--R & R--B \\
\hline
Fine
 & 2021 & Standard error & 0.046 & 0.021 & 0.020 & 0.023 & 0.034 & 0.018 \\
 &                       & Systematic     & 0.18  & 0.16  & 0.11  & 0.11  & 0.11  & 0.11  \\
 & 2022 & Standard error & 0.045 & 0.020 & 0.027 & 0.028 & 0.045 & 0.021 \\
 &                       & Systematic     & 0.19  & 0.19  & 0.29  & 0.29  & 0.23  & 0.22  \\
\hline
Coarse
 & 2021 & Standard error & 0.026 & 0.024 & 0.024 & 0.025 & 0.025 & 0.023 \\
 &                       & Systematic     & 0.06  & 0.05  & 0.02  & 0.02  & 0.04  & 0.04  \\
 & 2022 & Standard error & 0.062 & 0.051 & 0.042 & 0.043 & 0.039 & 0.033 \\
 &                       & Systematic     & 0.06  & 0.05  & 0.11  & 0.12  & 0.02  & 0.06  \\
\hline
\end{tabular}
\label{tab:uncertainties}
\end{table*}

The systematic term exceeds the statistical one in every case, typically
by a factor of three to five in the fine regime. The choice of pipeline
configuration, rather than the number of scans available, is therefore
the limiting factor on the accuracy of the derived \r{A}ngstr\"{o}m
exponent. Including the PSF correction inflates the systematic term by
roughly a factor of five in the fine regime, which reflects the outlying
behaviour of that configuration rather than a genuine increase in the
plausible range; the values excluding it are the ones adopted here.

The coarse regime carries a smaller systematic term than the fine
regime, since $\alpha$ is close to zero there and the configurations
differ correspondingly less in absolute terms. Its statistical
uncertainty in 2022 is nonetheless the largest in the table, reflecting
the 16 scans available in that regime and year. The standard error
understates the true uncertainty in that case, as those scans are drawn
from only a small number of nights and are therefore not independent
samples of the aerosol population.

The fact that two completely independent instruments --- a stellar
photometer operating at night and a sun photometer operating during
daytime --- both recover the same two aerosol populations at the same
site provides strong qualitative validation that the FRAM method is
capturing real atmospheric signal. A direct quantitative comparison
remains limited by the different wavelength pairs and the different
times of day at which the two instruments measure.

\subsection{CTAO-S0 and CTAO-S1, Chile}
\label{sec:results_ctas}

\begin{figure*}[h!]
\centering
\includegraphics[width=\textwidth]{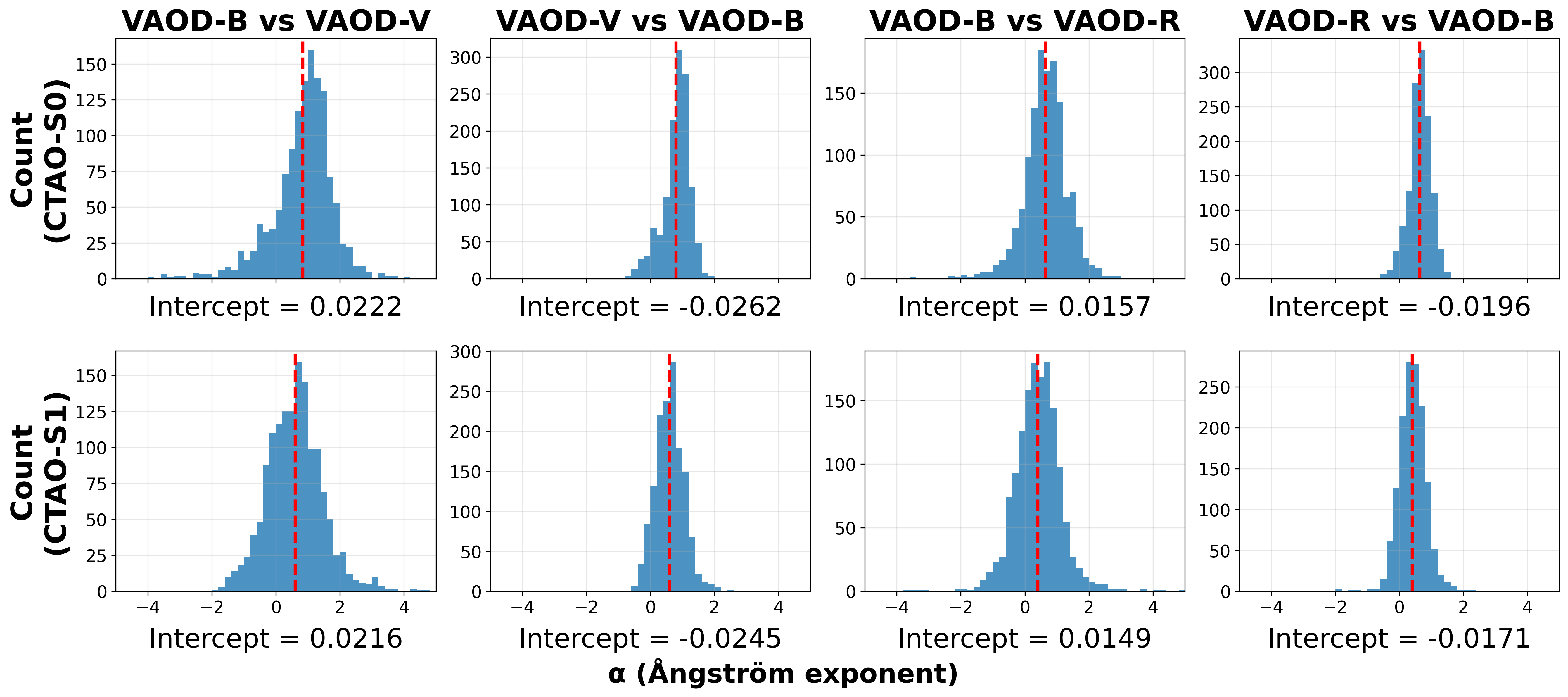}
\caption{\r{A}ngstr\"{o}m exponent ($\alpha$) distributions for CTAO-S0
(top row) and CTAO-S1 (bottom row) in the optimum aperture configuration,
for the four wavelength pairs involving the $B$ filter. CTAO-S1 is shown
with outlier scans removed. The dashed vertical line marks the mean
$\alpha$ and the fitted intercept is given beneath each panel. }
\label{fig:cta-south}
\end{figure*}

The \r{A}ngstr\"{o}m exponent distributions for CTAO-S0 and CTAO-S1
are shown in Figure~\ref{fig:cta-south}, after applying the intercept
correction described in
Section~\ref{sec:intercept}, with summary statistics for all
configurations in Tables~\ref{tab:stats_ctas0} and~\ref{tab:stats_ctas1}.
The two sites are located approximately 1.3~km apart at the Paranal
Observatory, with CTAO-S0 on an exposed ridge and CTAO-S1 in a more
sheltered valley. Unlike CTAO-N, the distributions at both southern
sites are unimodal, reflecting the more stable aerosol conditions at
Paranal. Since AERONET data were not available at Paranal for the period studied, external validation as performed for CTAO-N is not possible, and the analysis uses all available data rather than matched dates or a year-wise split.

\begin{table*}[htbp]
\centering
\caption{\r{A}ngstr\"{o}m exponent statistics --- CTAO-S0}
\begin{tabular}{lcccccccccccc}
\hline
Configuration & \multicolumn{2}{c}{B--V} & \multicolumn{2}{c}{V--B} &
\multicolumn{2}{c}{V--R} & \multicolumn{2}{c}{R--V} &
\multicolumn{2}{c}{B--R} & \multicolumn{2}{c}{R--B} \\
 & Mean & RMS & Mean & RMS & Mean & RMS & Mean & RMS &
   Mean & RMS & Mean & RMS \\
\hline
Optimum aperture & 0.84 & 1.27 & 0.80 & 0.92 & 0.40 & 0.55 & 0.40 & 0.59 & 0.65 & 0.94 & 0.64 & 0.73 \\
Aperture = 5     & 0.58 & 1.03 & 0.53 & 0.71 & 0.06 & 0.57 & 0.09 & 0.77 & 0.42 & 1.01 & 0.36 & 0.69 \\
Strat. aerosols  & 0.74 & 1.21 & 0.71 & 0.84 & 0.41 & 0.56 & 0.41 & 0.61 & 0.60 & 0.93 & 0.59 & 0.70 \\
No NLC           & 0.79 & 1.31 & 0.75 & 0.89 & 0.45 & 0.57 & 0.45 & 0.60 & 0.65 & 0.94 & 0.63 & 0.73 \\
PSF correction   & $-1.07$ & 1.99 & $-0.98$ & 1.74 & 0.41 & 1.58 & 0.41 & 1.73 & $-0.46$ & 1.36 & $-0.45$ & 1.36 \\
\hline
\end{tabular}
\label{tab:stats_ctas0}
\end{table*}

\begin{table*}[htbp]
\centering
\caption{\r{A}ngstr\"{o}m exponent statistics --- CTAO-S1}
\setlength{\tabcolsep}{4pt}
\begin{tabular}{lcccccccccccc}
\hline
Configuration & \multicolumn{2}{c}{B--V} & \multicolumn{2}{c}{V--B} &
\multicolumn{2}{c}{V--R} & \multicolumn{2}{c}{R--V} &
\multicolumn{2}{c}{B--R} & \multicolumn{2}{c}{R--B} \\
 & Mean & RMS & Mean & RMS & Mean & RMS & Mean & RMS &
   Mean & RMS & Mean & RMS \\
\hline
Optimum aperture & 0.60 & 1.07 & 0.60 & 0.75 & 0.13 & 0.55 & 0.14 & 0.60 & 0.41 & 0.84 & 0.41 & 0.61 \\
Aperture = 5     & 0.54 & 1.03 & 0.51 & 0.65 & $-0.13$ & 0.68 & $-0.08$ & 0.81 & 0.30 & 1.08 & 0.26 & 0.70 \\
Strat. aerosols  & 0.50 & 1.01 & 0.50 & 0.67 & 0.18 & 0.58 & 0.18 & 0.62 & 0.37 & 0.82 & 0.37 & 0.58 \\
No NLC           & 0.63 & 1.13 & 0.63 & 0.78 & 0.10 & 0.54 & 0.11 & 0.57 & 0.41 & 0.87 & 0.41 & 0.61 \\
PSF correction   & $-0.56$ & 1.47 & $-0.32$ & 1.08 & 0.36 & 1.27 & 0.40 & 1.42 & $-0.16$ & 1.13 & $-0.08$ & 0.97 \\
\hline
\end{tabular}
\label{tab:stats_ctas1}
\end{table*}

For the optimum aperture configuration, the B--V pair gives mean
$\alpha \approx 0.84$ at CTAO-S0 and $\alpha \approx 0.60$ at CTAO-S1,
and the B--R pair gives mean $\alpha \approx 0.65$ at CTAO-S0 and
$\alpha \approx 0.41$ at CTAO-S1. The V--R pair gives values close to
zero at both sites, consistent with its narrow wavelength baseline;
for this reason Figure~\ref{fig:cta-south} shows only the four pairs
involving the $B$ filter, which carry the wavelength leverage. The results at the two sites are in reasonable agreement
given that they are located 1.3~km apart and use instruments of similar
design, providing internal consistency validation of the method at the
southern sites. As shown in Tables~\ref{tab:stats_ctas0}
and~\ref{tab:stats_ctas1}, the distributions are consistent across four
of the five instrumental configurations at both sites. As at CTAO-N,
the PSF correction configuration is an exception, producing strongly
negative mean $\alpha$ values for B--V and B--R ($-1.07$ at CTAO-S0 and
$-0.56$ at CTAO-S1), confirming that the PSF correction as currently
implemented is not suitable for use.

\subsection{Sensitivity to Molecular Atmosphere Model}
\label{sec:mol_sensitivity}

\begin{figure*}[htbp]
    \centering
    \includegraphics[width=\textwidth]{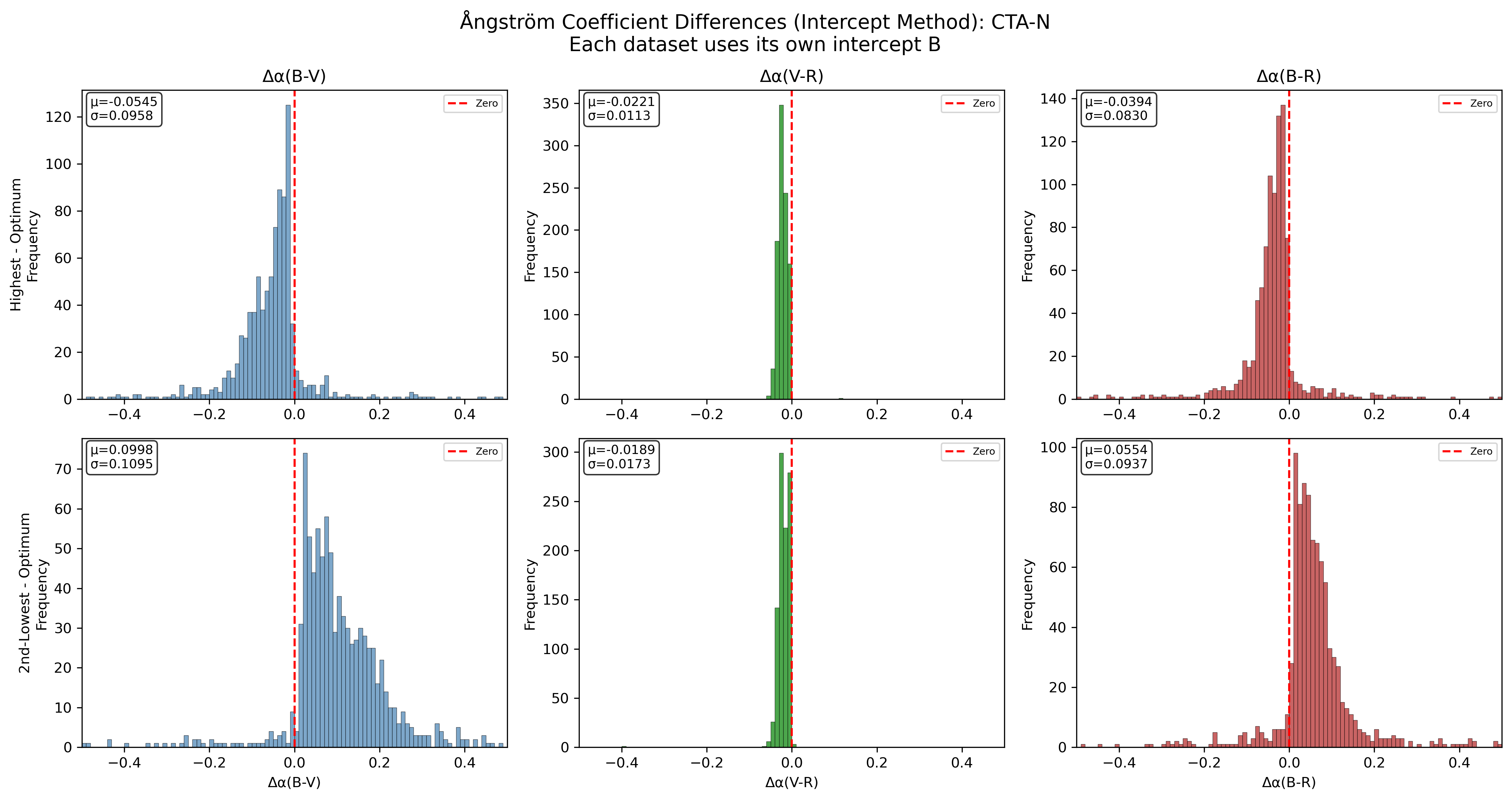}
    \caption{\r{A}ngstr\"{o}m exponent differences between the 
    highest and lowest ERA5 molecular atmosphere profiles and the 
    reference profile at CTAO-N, after applying the intercept 
    correction described in Section~\ref{sec:intercept}. Columns 
    show the three filter combinations: $\alpha$(B--V), $\alpha$(V--R), 
    and $\alpha$(B--R). Top row: highest molecular atmosphere minus 
    reference. Bottom row: lowest molecular atmosphere minus reference. 
    The vertical dashed line indicates zero difference.}
    \label{fig:angstrom_diff}
\end{figure*}

The molecular atmosphere used throughout this work is derived from 
a single reference ERA5 profile, from which the molecular extinction 
exponents in the magnitude equation (Equation~\ref{eq:mag}) are 
pre-computed. To assess the systematic uncertainty introduced by 
this choice, we compare VAOD and \r{A}ngstr\"{o}m exponent results 
derived using this reference profile against the highest and lowest 
available ERA5 profiles. It should be noted that none of these 
corresponds to the actual atmosphere on any given day of observation 
--- the mechanism for using the daily molecular atmosphere profile 
is not implemented in the pipeline, so all three represent extreme 
or mean cases that are in principle incorrect for any specific 
observation.

The mean VAOD offsets between the extreme profiles and the reference 
are well below the quoted measurement precision of the FRAM method 
of better than 0.008 optical depths \citep{Ebr2021}. For the highest 
profile, the offsets are $-0.0013$, $-0.0007$, and $-0.0005$ for 
the $B$, $V$, and $R$ filters respectively; for the lowest profile, 
they are $+0.0022$, $+0.0010$, and $+0.0006$. The decrease with 
increasing wavelength is consistent with the $\lambda^{-4}$ dependence 
of Rayleigh scattering. The choice of molecular atmosphere profile 
therefore introduces negligible uncertainty in the derived VAOD.

The impact on the \r{A}ngstr\"{o}m exponent is shown in 
Figure~\ref{fig:angstrom_diff}. The distributions are considerably 
broader than for VAOD --- this is a consequence of the intercept 
correction described in Section~\ref{sec:intercept}, which modifies 
the VAOD values before the ratio is computed. A small absolute 
change in VAOD translates into a much larger relative change in 
the ratio $\tau_x / (\tau_y - B)$, particularly at low aerosol 
optical depths where $\tau_y - B$ is small. Despite the broader 
distributions, the mean offsets remain small, confirming that the 
choice of molecular atmosphere profile does not significantly bias 
the derived \r{A}ngstr\"{o}m exponent.

\section{Conclusion}
\label{sec:conclusion}

This work extends the FRAM wide-field stellar photometry method from a 
single $B$-filter (440~nm) configuration using Tycho-2 to a simultaneous 
three-filter analysis in $B$, $V$, and $R$ (440, 540, and 620~nm) using 
Gaia~DR3 synthetic photometry. Gaia~DR3 was necessary because it 
provides photometry directly in the Johnson--Cousins system for all 
three filters, whereas Tycho-2 requires a passband transformation and 
has no $R$ photometry. The final pipeline configuration is summarised 
in Table~\ref{tab:final_config}. It uses an aperture radius of 3~pixels at CTAO-N 
and 2~pixels at CTAO-S0 and CTAO-S1, the same for all three filters.

The multi-filter VAOD measurements show a systematic filter-dependent 
ordering that is inconsistent with the physical expectation 
$\mathrm{VAOD}_B > \mathrm{VAOD}_V > \mathrm{VAOD}_R$. Although the 
$B$-filter VAOD agrees with the previously validated Tycho-2 result 
\citep{Ebr2021}, this does not establish which filter yields the 
correct value: the inconsistency could originate in the $B$ filter, in 
the $V$ and $R$ filters, or in all three. Four processing 
configurations, varying the aperture size, the aerosol airmass model, 
the non-linearity correction, and the use of a PSF aperture correction, 
did not resolve the ordering. We therefore introduced an intercept 
correction, which fits a linear relation between the VAOD of two 
filters using Orthogonal Distance Regression \citep{Boggs1990} and 
subtracts the non-zero intercept before computing the exponent. The PSF 
aperture correction increased the scatter and was not adopted. Replacing the reference ERA5 molecular profile with the 
highest and lowest available profiles changes the mean VAOD by at most 
0.0022, below the FRAM measurement precision of 0.008, so the molecular 
atmosphere introduces negligible uncertainty.

The intercept correction is possible because the \r{A}ngstr\"{o}m 
exponent is approximately constant within each aerosol population. At 
CTAO-S0 and CTAO-S1 the aerosol conditions are relatively stable, while 
at CTAO-N the aerosol population separates into two regimes by VAOD. 
For the CTAO-N low aerosol loading population the scatter of the VAOD 
pairs about the fitted line (RMS of 0.002--0.006 across the three 
filter pairs and both years) is smaller than the quoted FRAM 
measurement precision of 0.008 optical depths \citep{Ebr2021}, which 
implies that the exponent is nearly constant within this population. 
The intercepts are similar at the three sites: 0.022--0.029 for B--V, 
$-0.017$ to $-0.032$ for R--B, and $-0.006$ for V--R at every site 
except CTAO-N in 2022 ($-0.002$). This suggests a common systematic 
origin in the instruments or the processing rather than site-specific 
atmospheric conditions, but the origin remains unidentified. The 
exponent derived directly from the slope of the VAOD--VAOD relation 
agrees with the intercept-corrected mean to within 0.15 for all six 
filter pairs in both years.

At the two southern Paranal sites, which are 1.3~km apart and use 
instruments of similar design, the exponent distributions are unimodal. 
For the optimum aperture configuration the mean exponent is 
$\alpha \approx 0.84$ (B--V) and $0.65$ (B--R) at CTAO-S0, and $0.60$ 
and $0.41$ at CTAO-S1. At CTAO-N the B--V and B--R distributions are 
bimodal, as \citet{Schmuckermaier2023} found at the same site using 
AERONET data. The coarse regime, which corresponds to Calima dust 
intrusions, has a mean B--V exponent of 0.18, and the fine regime, 
which corresponds to the clean background, has 1.33.

We compared the CTAO-N results with AERONET Level~2.0 sun-photometer 
measurements from the same site, restricted to the 300 calendar dates 
on which both instruments obtained data and analysed separately for 
2021 and 2022, because the seasonal coverage of the common dates 
differs between the two years. The FRAM VAOD$_B$ thresholds of 0.04 and 
0.1 were taken from the AERONET AOD distribution at 440~nm. The FRAM 
coarse regime agrees with the AERONET coarse population (B--V exponent 
of 0.18 against 0.21), whereas the FRAM fine regime is lower than the 
AERONET value (1.33 against 1.96), though within the range found by 
\citet{Schmuckermaier2023} (1.45 $\pm$ 0.46). In the fine regime the 
systematic uncertainty from the choice of processing configuration 
(about 0.2 for B--V) exceeds the statistical uncertainty (about 0.05). 
That two independent instruments, one operating at night and one during 
the day, recover the same two populations at the same site provides 
qualitative support for the method.

We recommend the B--V and B--R pairs for determining the 
\r{A}ngstr\"{o}m exponent, since they separate the two aerosol regimes 
most clearly. The V--R pair is the least robust, because its 80~nm 
separation amplifies changes in the fitted slope and in the intercept 
subtraction by a factor of 7.2, compared with 4.9 for B--V and 2.9 for 
B--R. Future work will aim to identify the origin of the 
filter-dependent VAOD ordering, improve the PSF aperture correction, and 
extend the analysis to further observing seasons, in particular the 
late-summer months at CTAO-N that are missing from the common dates in 
2022.

\begin{acknowledgments}
This work was supported by the Ministry of Education, Youth and 
Sports LM2023047 and EU/MEYS CZ.02.01.01/00/22\_008/0004632. 
This work was conducted in the context of the CTA Consortium. 
We thank the AERONET team and, specifically, Dr.~\'{A}frica 
Barreto and AEMET staff for establishing and maintaining the 
Roque de los Muchachos AERONET site used in this investigation.
\end{acknowledgments}

\bibliography{sample701}{}
\bibliographystyle{aasjournalv7}

\end{document}